\documentclass[aps,prd,showpacs,preprintnumbers]{revtex4}
\usepackage{amsmath,epsfig}
\usepackage{subfigure}
\usepackage{amssymb,amsfonts}
\usepackage{latexsym}
\usepackage{epsfig}
\usepackage{amssymb}
 \usepackage{bm}
\newcommand{\be}{\begin{eqnarray}}
\newcommand{\ee}{\end{eqnarray}}

\newcommand{\bea}{\begin{eqnarray}}
\newcommand{\eea}{\end{eqnarray}}

\newcommand{\non}{\nonumber\\}

\newcommand{\eq}{\begin{equation}}
\newcommand{\eeq}{\end{equation}}
\newcommand{\eqn}{\begin{eqnarray}}
\newcommand{\eeqn}{\end{eqnarray}}
\begin{document}
\title{Quark matter under rotation in the NJL model with vector interaction}
\author{Xinyang Wang$^{a,b}$}
\email{wangxinyang@ihep.ac.cn}
\author{Minghua Wei$^{a,c}$}
\email{weimh@ihep.ac.cn}
\author{Zhibin Li$^{a,c}$}
\email{lizb@ihep.ac.cn}
\author{Mei Huang$^{c,a,d}$}
\email{huangmei@ucas.ac.cn}
\affiliation{$^{a}$  Institute of High Energy Physics, Chinese Academy of Sciences, Beijing 100049, P.R. China}
\affiliation{$^{b}$  Department of Physics, Jiangsu University, Zhejiang 212013, P.R. China}
\affiliation{$^{c}$ School of Nuclear Science and Technology, University of Chinese Academy of Sciences, Beijing 100049, P.R. China}
\affiliation{$^{d}$ Theoretical Physics Center for Science Facilities, Chinese Academy of Sciences, Beijing 100049, P.R. China}

\begin{abstract}
We study the chiral phase transition of quark matter under rotation in two-flavor Nambu--Jona-Lasinio (NJL) model. It is found that, in the rotating frame, the angular velocity plays the similar role as the baryon chemical potential and suppresses the chiral condensate, thus the chiral phase transition shows a critical end point not only in the temperature-chemical potential $T-\mu$ plane, but also in the temperature-angular momentum $T-\omega$ plane. One interesting observation is that in the $T-\mu$ plane, the presence of the angular momentum only shifts down the critical temperature $T^E$ of the CEP and does not shift the critical chemical potential $\mu^E$, and in the $T-\omega$ plane, the increase of the chemical potential only shift down the critical temperature $T^E$ and does not change the critical angular momentum $\omega^E$. The phase structure in the $T-\mu$ plane is sensitive to the coupling strength in the vector channel, while the phase structure in $T-\omega$ plane is not. It is also observed that the rotating angular velocity suppresses the kurtosis of the baryon number fluctuations, while it enhances the pressure density, energy density, the specific heat and the sound velocity.
\end{abstract}
\maketitle

\section{Introduction}

The phase transitions and phase structure of Quantum Chromodynamics (QCD) at finite temperature/density and other extreme conditions are the main topics of relativistic heavy ion collisions. The properties of QCD matter and its equation of state is also highly related to the evolution of the early universe and mass-radius relation of the compact stars. Recently, lots of interests are attracted to investigate magnetized and fast rotating QCD matter because strong magnetic field and
large angular momentum can be created through non-central heavy ion collisions. In the past decade, lots of studies focus on QCD matter under strong magnetic fields and many interesting phenomena have been discussed, e.g. the chiral magnetic effect (CME)~\cite{Kharzeev:2007tn,Kharzeev:2007jp,Fukushima:2008xe}, the magnetic catalysis~\cite{Klevansky:1989vi,Klimenko:1990rh,Gusynin:1995nb} and inverse magnetic catalysis~\cite{Bali:20111213} effect, and the vacuum superconductivity~\cite{Chernodub:2010qx,Chernodub:2011mc}. However, studies on properties of fast rotating QCD matter are still relatively few.

Rotating matter exists in many physical environments, e.g. the rapidly spinning neutron stars in astrophysics, the trapped nonrelativistic bosonic cold atoms in condensed matter physics and non-central heavy-ion collisions in high energy nuclear physics. For the off-central heavy ion collisions, the two colliding nuclei carry a total momentum $J\propto b\sqrt{s_{NN}}$. Here, b is the impact parameter and the beam energy $\sqrt{s_{NN}}$ is nucleon-nucleon center-of-mass energy. After the collision, most of the angular momentum is carried away by the spectators, and there still remains a nonzero angular momentum in the range of $10^3 \hbar-10^5 \hbar$ with local angular velocity in the range of $0.01 {\rm GeV}-0.1{\rm GeV}$ in the created hot QCD matter~\cite{Jiang:2016woz,Deng:2016gyh}. Some interesting physical phenomena were found in rotating QCD matter (i.e., chiral vortical effect~\cite{Kharzeev:2007tn,Son:2009tf,Kharzeev:2010gr} and chiral vortical wave~\cite{Jiang:2015cva}). The chiral vortical effect and chiral vortical wave play analogous roles to the chiral magnetic effect~\cite{Kharzeev:2007tn,Kharzeev:2007jp} and the chiral magnetic wave~\cite{Kharzeev:2010gd,Burnier:2011bf}, which were found in magnetized matter (matter in strong magnetic fields).

There are already lots of studies on QCD phase structure and properties of QCD matter under strong magnetic fields (for review see Ref.~\cite{Andersen:2014xxa}). Comparing with that, studies on the phase diagram of fast rotating QCD matter are quite limited, e.g. see references in~\cite{Chen:2015hfc,Jiang:2016wvv,Ebihara:2016fwa,Chernodub:2016kxh,Chernodub:2017ref}. It was observed in~\cite{Jiang:2016wvv} that there is a generic suppression effect on both quark-antiquark and diquark pairing states with zero angular momentum, and there is another critical end point in the temperature-rotation parameter space. In this work, we investigate the QCD phase diagram of fast rotating quark matter at finite temperature and
density in the two-flavor NJL model with vector interaction. In recent studies, it is pointed out in~\cite{Chen:2015hfc,Ebihara:2016fwa,Chernodub:2016kxh,Chernodub:2017ref} that the boundary effect is very important in rotating system, since all the research are considering the leading order of angular velocity in Lagrangian expansion, strictly speaking, this is only true when  the angular velocity is much smaller than the inverse of system's size and thus can ignore the finite volume boundary effect. So in this paper,  we just ignore the boundary effect and give a qualitative result. We show a 3D phase structure in the $T-\mu-\omega$ frame, and carefully investigate the influence of the angular velocity $\omega$ on the CEP in the $T-\mu$ plane. It is interesting to notice that the presence of the angular momentum only shifts down the critical temperature $T^E$ of the CEP and does not shift the critical chemical potential $\mu^E$. We also find that with fixed chemical potential, the chiral phase transition at high angular momentum is of first order, and there is another CEP shows up in the temperature-angular momentum $T-\omega$ plane. Similarly we find that in the $T-\omega$ plane, the increase of the chemical potential only shift down the critical temperature $T^E$ and does not change the critical angular momentum $\omega^E$. We also evaluate how the vector interaction will affect the phase diagram. The numerical result shows that the influence of vector interaction on the chiral phase transition in the T-$\omega$ plane is much less sensitive comparing with the the chiral phase transition in the T-$\mu$ plane.

This paper is organized as following: in the next section, we give a general expression of the two-flavor NJL model including vector interaction in rotating frame, and then derive thermodynamical potential and the gap equations for the chiral condensate. In Sec.~\ref{Numerical} we show our numerical results and analysis on the CEP and influence from the vector interaction. An experimental relative quantity which is the kurtosis of baryon number fluctuation has been evaluated with different angular momentum. Several thermodynamic quantities are also computed.  Finally, the discussion and conclusion is given in Sec.~\ref{Conclusion}.

\section{Formalism}
The Lagrangian of the two-flavor NJL model with vector interaction in the rotating frame is given by~\cite{Bernard:1988db}:
\eqn
\mathcal{L} = \bar{\psi}[i\gamma^{\mu}(\partial_{\mu}+\Gamma_{\mu})-m]\psi+G_S[(\bar{\psi}\psi)^2+(\bar{\psi}i\gamma_5\vec{\tau}\psi)^2]-G_V[(\bar{\psi}\gamma_{\mu}\psi)^2+(\bar{\psi}\gamma_{\mu}\gamma_5\psi)^2].
\eeqn
Here, we consider the system with a constant angular velocity $\vec{\omega}$ along z-axis and $\vec{v} =\vec{\omega}\times\vec{x}$ is the local veclocity of this rotating frame. $m$ is the current quark mass, G$_S$ and G$_V$ are the coupling constants in the scalar and vector channels, respectively. The spinor connection is given by $\Gamma_\mu=\frac{1}{4}\times\frac{1}{2}[\gamma^a,\gamma ^b] \, \Gamma_{ab\mu}$. Here, $\Gamma_{ab\mu}=\eta_{ac}(e^c_{\ \sigma} G^\sigma_{\ \mu\nu}e_b^{\ \nu}-e_b^{\ \nu}\partial_\mu e^c_{\ \nu})$, $G^\sigma_{\ \mu\nu}$ is the affine connection determined by $g^{\mu\nu}$. Following Ref.~\cite{Jiang:2016wvv}, we expand the Lagrangian up to the leading order of $\omega$ and choose $e^{a}_{\ \mu}=\delta^a_{\ \mu}+  \delta^a_{\ i}\delta^0_{\ \mu} \, v_i$ and $e_{a}^{\ \mu}=\delta_a^{\ \mu} -  \delta_a^{\ 0}\delta_i^{\ \mu} \, v_i$.
Following the
derivation in \cite{Jiang:2016wvv} and \cite{Buballa:2003qv}, the Lagrangian with vector interaction in the mean field approximation is given by
\eq
\mathcal{L}=\bar{\psi}[i\bar{\gamma^{\mu}}(\partial_{\mu}+\gamma^{0}\omega \hat{J_{z}})-M]\psi+(\tilde{\mu}-\mu)\psi^{\dag}\psi-\frac{(M-m)^{2}}{4G_{S}}+\frac{(\mu-\tilde{\mu})^{2}}{4G_{V}}.
\eeq
Where $J_{z}$ is the third direction of total angular momentum, the effective quark chemical potential is defined as $\tilde{\mu}=\mu-2G_V \left<\psi^{\dagger}\psi\right>$, and the constituent quark mass in the mean-field approximation is given by $M = m - 2G_S\left<\bar\psi\psi\right>$.
The general grand potential is given by
\eq
\Omega(T,\mu;M,\tilde{\mu},\omega)=\Omega_{M}(T,\mu;M,\tilde{\mu},\omega)+\int d^3 \mathbf{r}~ \bigg\{\frac{(M-m)^{2}}{4G_{S}}-\frac{(\mu-\tilde{\mu})^{2}}{4G_{V}}\bigg\}.
\eeq
Using the standard method from the textbook~\cite{Kapusta}, we could get
\eqn
&&\Omega_{M}(T,\mu;M,\tilde{\mu},\omega)=\int d^3 \mathbf{r}~ \left\{-\frac{T}{32\pi^2}\sum_{N}\sum_n\int d k_t^2 \int d k_z[J_n(k_t r)^2+J_{n+1}(k_t r)^2]\text{Tr}\ln D\right\}\non
&=&\int d^3 \mathbf{r}~ \left\{-\frac{T N_c N_f}{16\pi^2}\sum_{N}\sum_n\int d k_t^2 \int d k_z[J_n(k_t r)^2+J_{n+1}(k_t r)^2]
\ln\{\beta^{2}[(\omega_{N}+i\tilde{\mu}+i(n+\frac{1}{2})\omega)^{2}+E^{2}_{k}]\}\right\},
\eeqn
with
\eq
D=-i\beta[(-i\omega_{N}+\tilde{\mu}+(n+\frac{1}{2})\omega)-\gamma^{0}\vec{\gamma}\cdot\vec{k}-M\gamma^{0}].
\eeq
Here, $k_z$ the momentum in the z-direction and $k_t$ the transverse momentum, $E_k=\sqrt{k_z^2+k_t^2+M^2}$. $\beta = 1/T$ and the Matsubara frequency $\omega_{N}=(2N+1)\pi T$, $\mathbf{r}$ is the location from the center of rotation, $J_n(x)$ is the first kind nth Bessel functions with $ n = 0, \pm1,...$ the z-angular-momentum quantum number.

Using the following relations,
\eqn
&&2\sum_{N}\ln\{\beta^{2}[(\omega_{N}+i\tilde{\mu}+i(n+\frac{1}{2})\omega)^{2}+E^{2}_{k}]\}\non
&=&\sum\limits_{N}
\bigg\{\ln[\beta^{2}(\omega^{2}_{N}+(E_{k}-\tilde{\mu}-(n+\frac{1}{2})\omega)^{2})]
+\ln[\beta^{2}(\omega^{2}_{N}+(E_{k}+\tilde{\mu}+(n+\frac{1}{2})\omega)^{2})]\bigg\}\non
&=&\beta E_{k}+\ln(1+e^{-\beta(E_{k}-(n+\frac{1}{2})\omega-\tilde{\mu})})+\ln(1+e^{-\beta(E_{k}+(n+\frac{1}{2})\omega+\tilde{\mu})})\non
&=&-\beta E_{k}+\ln(1+e^{\beta(E_{k}-(n+\frac{1}{2})\omega-\tilde{\mu})})+\ln(1+e^{\beta(E_{k}+(n+\frac{1}{2})\omega+\tilde{\mu})}),
\eeqn
we get
\eqn
\Omega_{M}(T,\mu;M,\tilde{\mu},\omega)&=&\int d^3 \mathbf{r}~ \bigg\{ -\frac{N_c N_f}{16\pi^2} T \sum_n\int dk_t^2 \int dk_z[J_n(k_t r)^2+J_{n+1}(k_t r)^2]\left[ \ln(1+e^{(E_k-(n+\frac{1}{2})\omega-\tilde{\mu})/T})\right.\non
& & + \left. \ln(1+e^{-(E_k-(n+\frac{1}{2})\omega-\tilde{\mu})/T})+\ln(1+e^{-(E_k+(n+\frac{1}{2})\omega+\tilde{\mu})/T})+ \ln(1+ e^{(E_k+(n+\frac{1}{2})\omega+\tilde{\mu})/T})\right]\bigg\}.\non
\eeqn
Notice here a factor of 2 is taking account of particles and antiparticles. Then the general grand potential function becomes
\eqn
\label{omg}
\Omega(T,\mu;M,\tilde{\mu},\omega) &=&\int d^3 \mathbf{r}~ \bigg\{\frac{(M-m)^2}{4G_S} -\frac{(\mu-\tilde{\mu})^2}{4G_V} \nonumber \\
     & & -\frac{N_c N_f}{16\pi^2} T \sum_n\int dk_t^2 \int dk_z[J_n(k_t r)^2+J_{n+1}(k_t r)^2]\left[ \ln(1+e^{(E_k-(n+\frac{1}{2})\omega-\tilde{\mu})/T})\right.\non
& & + \left. \ln(1+e^{-(E_k-(n+\frac{1}{2})\omega-\tilde{\mu})/T})+\ln(1+e^{-(E_k+(n+\frac{1}{2})\omega+\tilde{\mu})/T})+ \ln(1+ e^{(E_k+(n+\frac{1}{2})\omega+\tilde{\mu})/T})\right]\bigg\}.\non
\eeqn
In order to find the stationary points of $\Omega$ with respect to M and  $\tilde{\mu}$, we need to solve the following gap equations,
\eqn
\frac{\partial \Omega}{\partial M} = 0,~~\frac{\partial \Omega}{\partial \tilde{\mu}} = 0,
\eeqn
with the following constraint,
\eqn
\frac{\partial^2 \Omega}{\partial M^2} > 0.
\eeqn
The gap equations take the following forms:
\begin{subequations}
\eqn
0&=&\int d^3 \mathbf{r}\left\{ \frac{M-m}{2G_S}-\frac{N_c N_f}{8\pi^2}\sum_n\int dk_t^2 \int dk_z[J_n(k_t r)^2+J_{n+1}(k_t r)^2]\frac{ M \sinh\left(\frac{E_k}{T}\right)}{E_k\left[\cosh\left(\frac{E_k}{T}\right)+\cosh\left(\frac{\tilde{\mu}+(n+\frac{1}{2})\omega}{T}\right)\right]}\right\},\non
\eeqn
\eqn
0&=&\int d^3\mathbf{r} \left\{  \frac{\mu-\tilde{\mu}}{2G_V}-\frac{N_c N_f}{8\pi^2}\sum_n\int dk_t^2 \int dk_z[J_n(k_t r)^2+J_{n+1}(k_t r)^2]\frac{ \sinh\left(\frac{\tilde{\mu}+(n+\frac{1}{2})\omega}{T}\right)}{\cosh\left(\frac{E_k}{T}\right)+\cosh\left(\frac{\tilde{\mu}+(n+\frac{1}{2})\omega}{T}\right)}\right\},
\eeqn
with constraint
\eqn
&&\int d^3 \mathbf{r}\left\{ \frac{1}{2G_S}-\frac{N_c N_f}{8\pi^2}\sum_n\int dk_t^2 \int dk_z\frac{[J_n(k_t r)^2+J_{n+1}(k_t r)^2]}{T E_k^{3}\left(\cosh(\frac{E_k}{T})+\cosh(\frac{\tilde{\mu}+2(n+1)\omega}{T})\right)^2}\right.\non
&\times&\left[M^2E_k\cosh\left(\frac{E_k}{T}\right)\cosh\left(\frac{\tilde{\mu}+(n+1/2)\omega}{T}\right)+k^2T\sinh\left(\frac{E_k}{T}\right)\left(\cosh\left(\frac{E_k}{T}\right)+\cosh\left(\frac{\tilde{\mu}+(n+1/2)\omega}{T}\right)\right)\right.\non
&-&\left.\left.M^2E_k\sinh^2\left(\frac{E_k}{T}\right)+M^2E_k\cosh^2\left(\frac{E_k}{T}\right)\right]\right\}>0.
\eeqn
\end{subequations}

\section{Numerical results}
\label{Numerical}

We consider the two-flavor case $N_f=2$ and take $N_c=3$. For numerical calculations we choose one set of parameters for the current quark mass $m$ and
the coupling constant in the scalar channel $G_S$
\eqn
\hspace{0.3cm} m=5.5~\text{MeV},\hspace{0.3cm} G_S=5.04\times10^{-6}~\text{MeV}^{-2}, \Lambda=651~\text{MeV}
\eeqn
by fitting the pion mass and pion weak decay constant in the vacuum as in Ref.~\cite{Klevansky:1992qe}. In the vacuum the coupling constant $G_V$ in the vector channel should be determined by the vector mesons as shown in Refs.~\cite{Bernard:1995hm}. However, the coupling constants may change in the medium, for example, it was shown in Ref.~\cite{Schafer:1994nv,Schafer:1996wv} that the coupling constant in the vector as well as axial-vector channels will change sign due to the instanton-anti-instanton pairing effect at high temperature~\cite{Schafer:1994nv} when chiral symmetry restores. In Ref.~\cite{Yu:2014sla}, the effect of negative $G_A$ above $T_c$ has been analyzed. In this work, we treat the coupling constant $G_V$ as a free parameter. In the original NJL model with only scalar interaction, the predicted critical end point (CEP) is located at a rather low temperature region. It was shown in Ref.~\cite{Bratovic:2012qs,Hell:2012da} that changing the value of the coupling constant in the flavor singlet vector interaction $G_V$ will shift the location of CEP, and with the increasing of positive $G_V$, the CEP will disappear in the $T-\mu$ plane. It has been shown in Ref.~\cite{Li:2018ygx}, a negative vector coupling constant will raise the location of CEP which may have a better agreement with the experiment measurement of the baryon number fluctuations~\cite{Luo:2017faz}. Therefore, we treat the coupling constant $G_V$ as a free parameter to shift the location of the CEP, and we take $G_V=0$, $G_V=0.67 G_S$ and $G_V=-0.5 G_S$ in our calculations for comparison.
Also by following Ref.~\cite{Jiang:2016wvv} we pick up a particular value of transverse radial coordinate $r = 0.1$ GeV$^{-1}$ in our calculation, which is a typical radius in heavy ion collisions.

\subsection{Phase Diagram in the $T-\mu$ and $T-\omega$ plane}
\begin{figure}[t!]
\includegraphics[width=6.5cm]{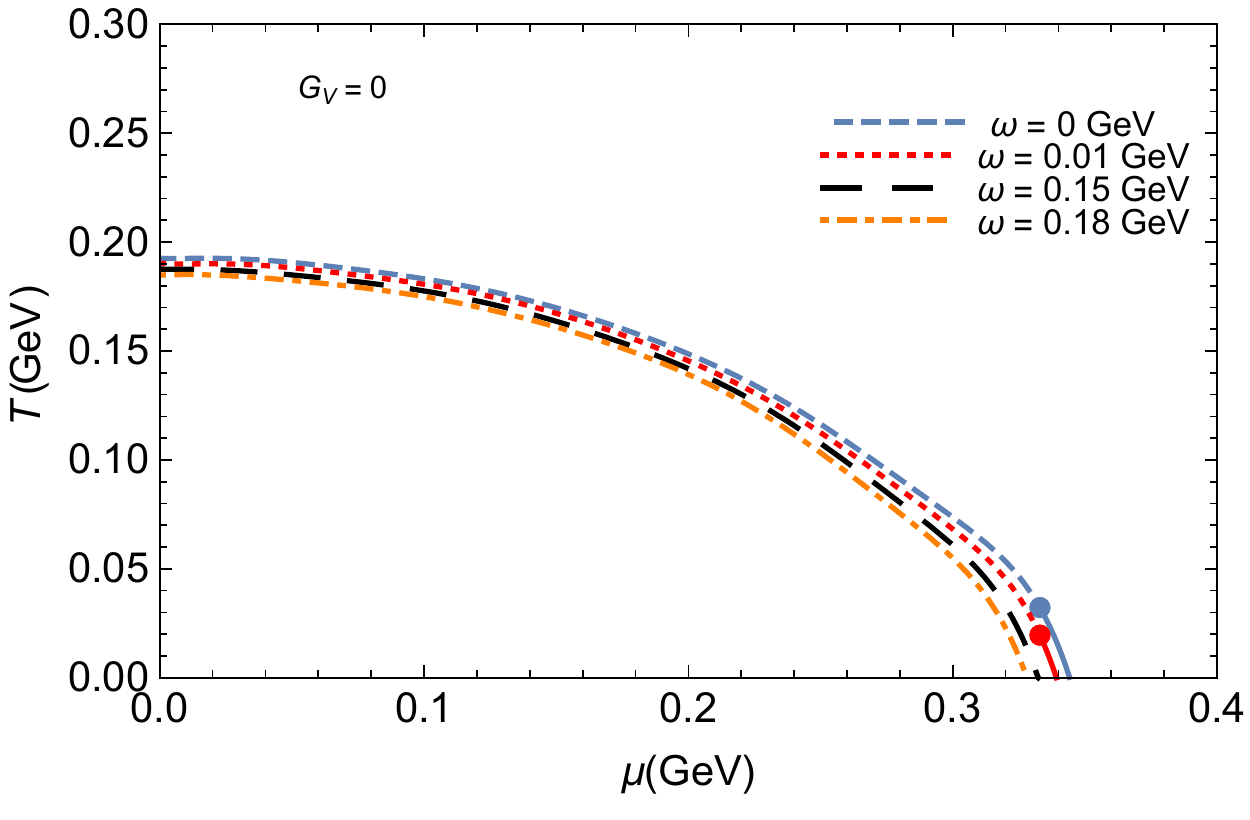}
\includegraphics[width=6.5cm]{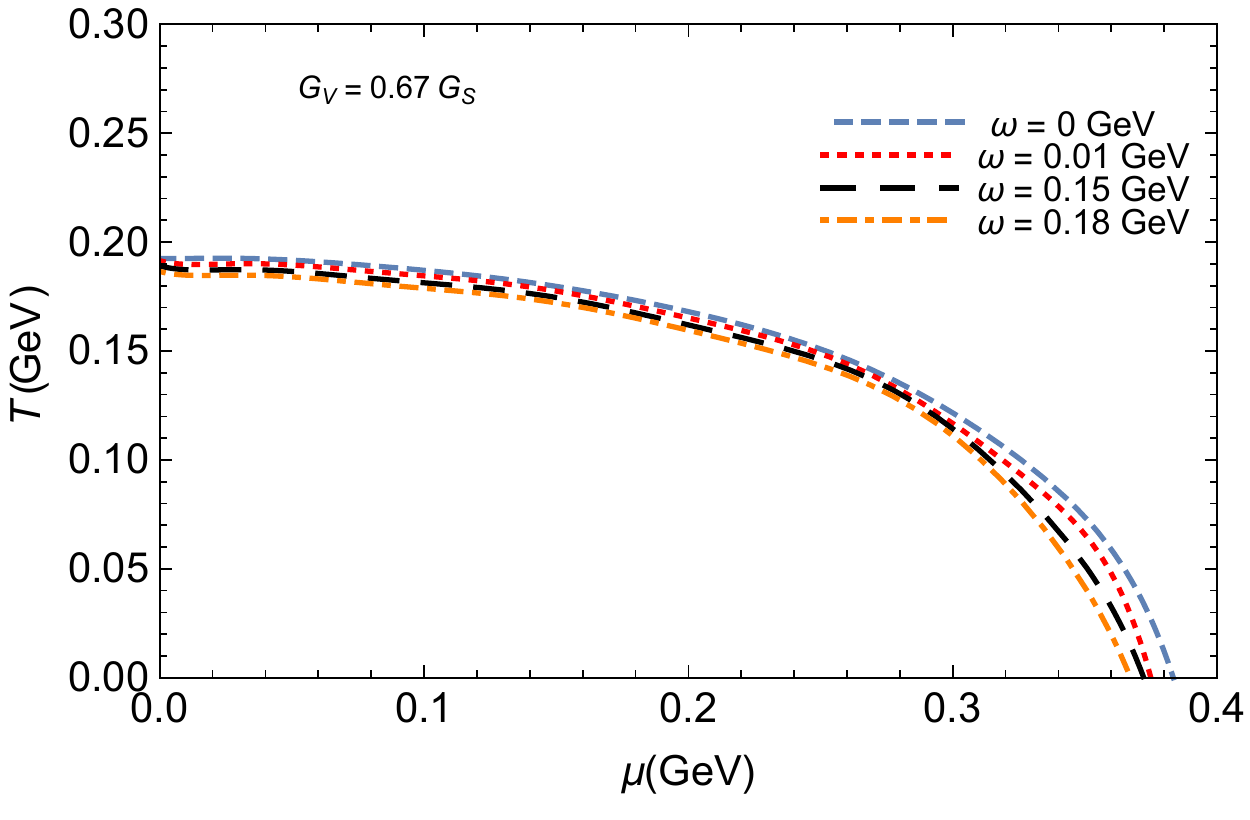}
\includegraphics[width=6.5cm]{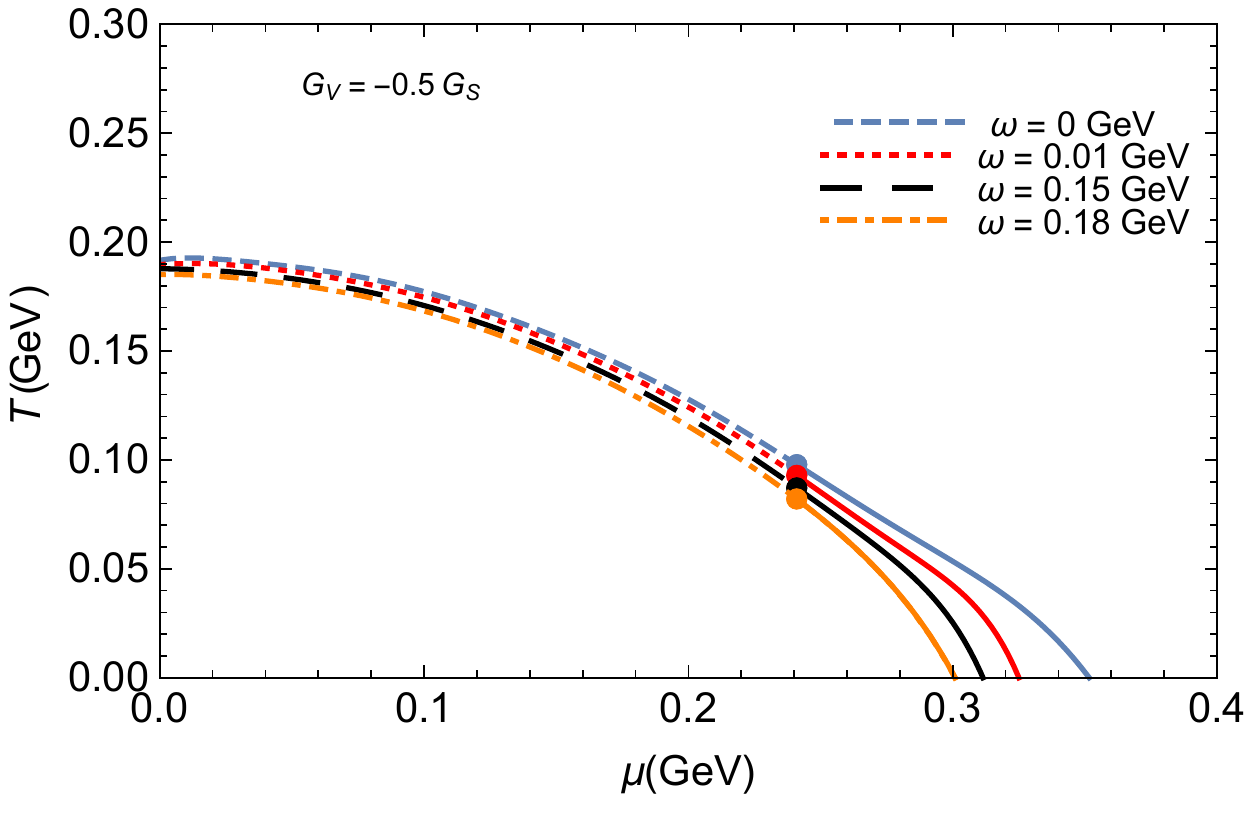}
\caption{The phase diagram in the T-$\mu$ plane with different $\omega$ with the coupling constant in the vector channel takes the value of G$_V$ = 0, 0.67 G$_S$ and -0.5 G$_S$ from top left to bottom.}
\label{fig:fig1}
\end{figure}

\begin{figure}[t!]
\includegraphics[width=6.5cm]{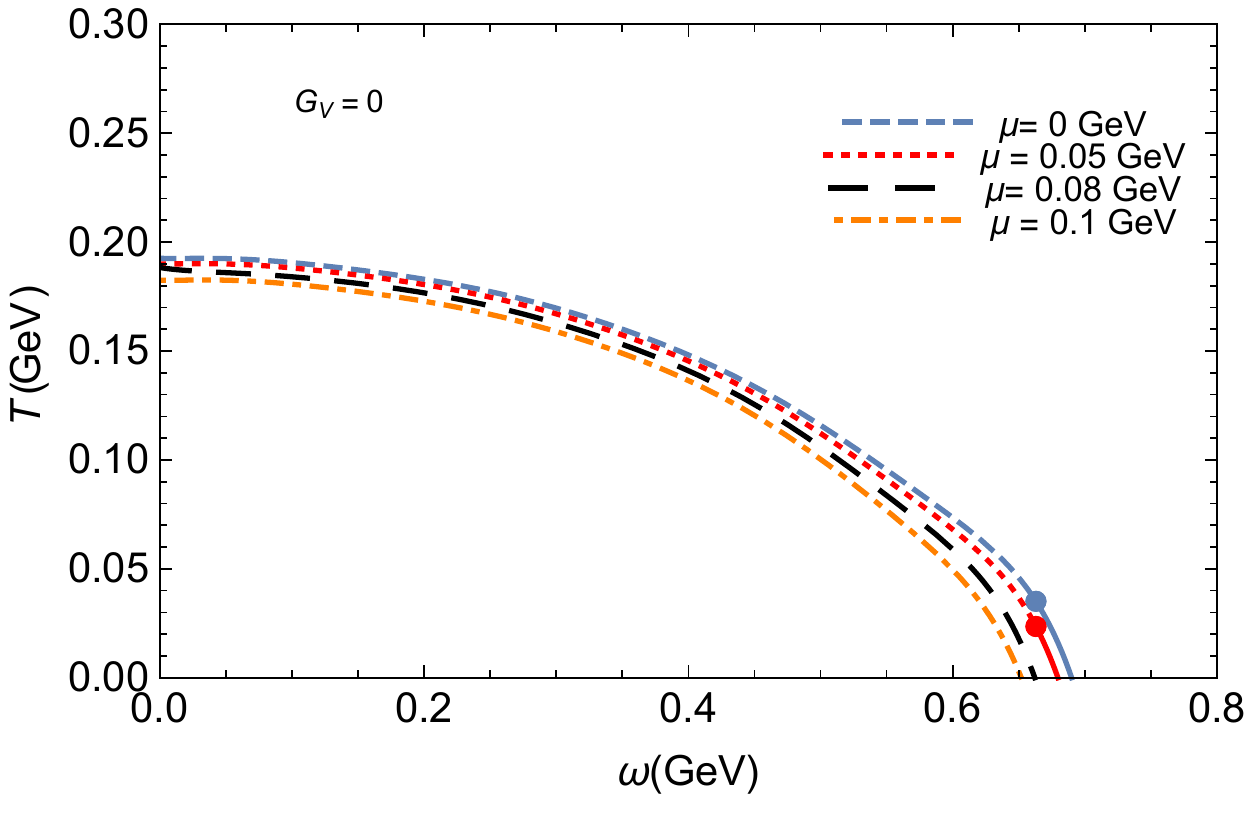}
\includegraphics[width=6.5cm]{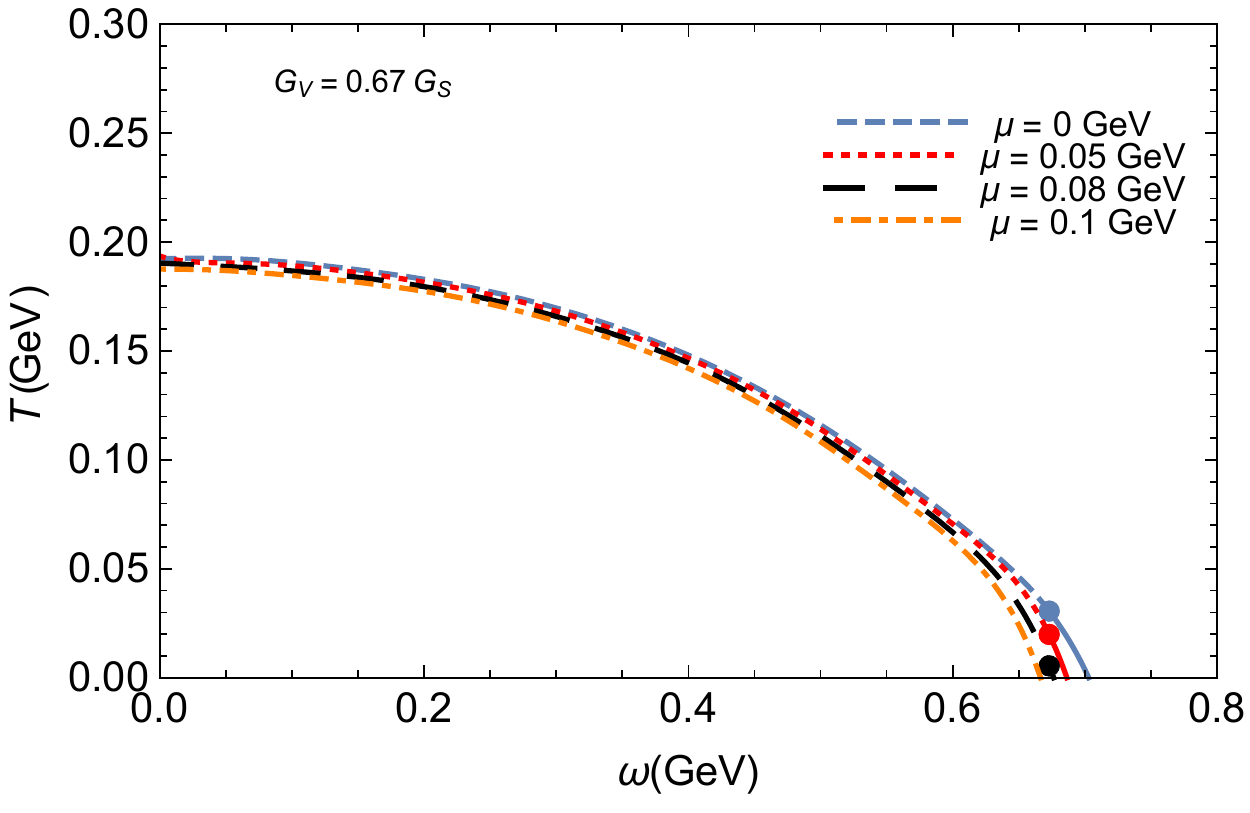}
\includegraphics[width=6.5cm]{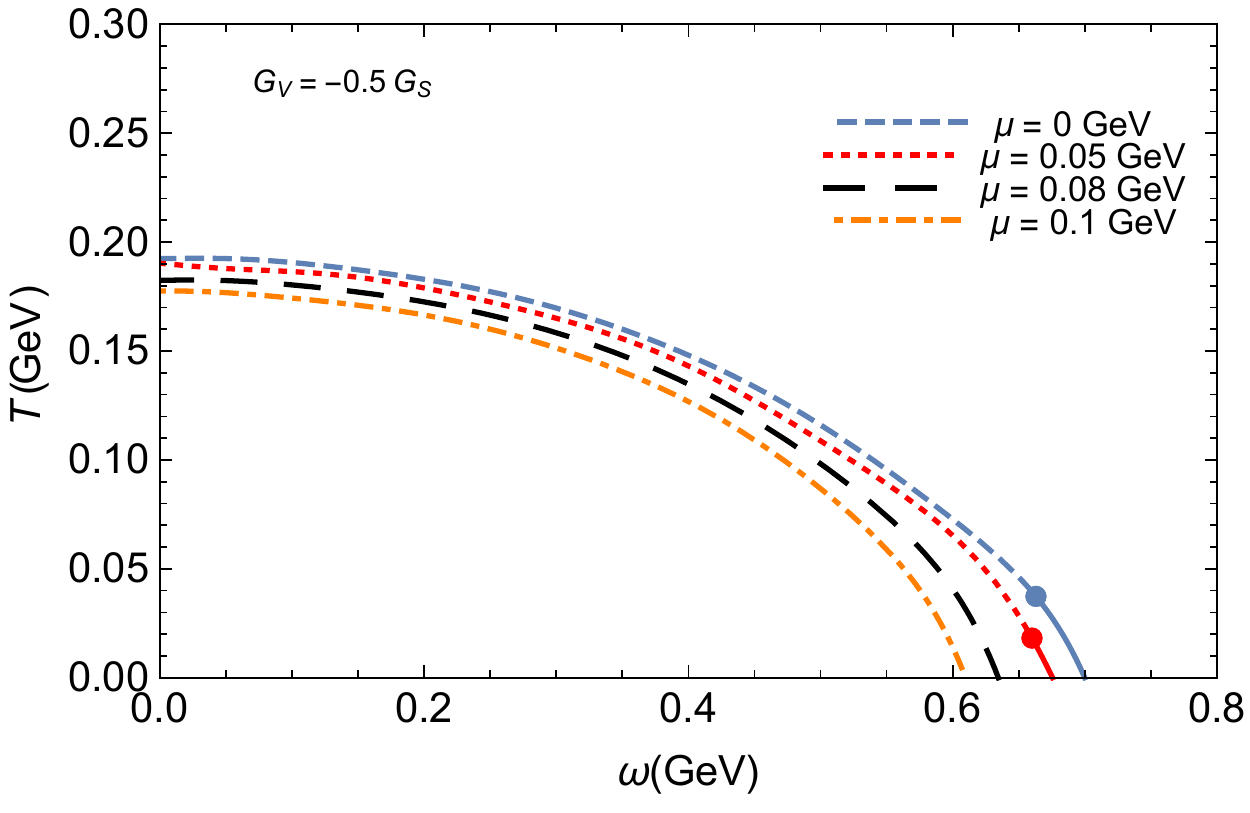}
\caption{The phase diagram in the T-$\omega$ plane with different chemical potentials $\mu$. The vector couplings are G$_V$ = 0, 0.67 G$_S$ and -0.5 G$_S$ from top left to bottom.}
\label{fig:fig2}
\end{figure}

By solving the gap equations and finding the minima of the thermodynamical potential, we give the phase diagram of $T-\mu$ and $T-\omega$ in Figs.~\ref{fig:fig1} and Figs.~\ref{fig:fig2}, respectively. The dashed, dotted and dash-dotted lines represent crossover, the solid lines represent the phase transitions and the big dots are the CEP points under each conditions.

Fig.~\ref{fig:fig1} shows the $T-\mu$ phase diagram with different angular velocities $\omega=0,0.01,0.15,0.18$ for G$_V$ = $0$, $0.67$G$_S$ and $-0.5$G$_S$, respectively. The corresponding locations of the CEP are listed in Table~\ref{table:table1}. For the case of G$_V$ = 0 and $\omega=0$, the CEP is located at the tail of the phase boundary at $T^E=0.032 {\rm GeV}, \mu^E=0.333 {\rm GeV}$. A positive coupling constant in the vector channel shifts away the CEP from the phase diagram, and a negative coupling constant in the vector channel shifts the CEP to the left part of the phase boundary with a higher critical temperature and a lower critical baryon chemical potential. It is noticed that with fixed coupling constant in the vector channel, the increase of the angular momentum does not change the phase boundary so much in the small baryon density region but changes the phase boundary in the large baryon density region. For repulsive interaction in the vector channel G$_V$=-0.5G$_S$, the angular velocity has larger effect on the phase boundary in the baryon density region higher than the critical baryon density $\mu>\mu^E$. Another interesting observation is that the angular velocity only shifts down the critical temperature $T^E$ and does not change the critical chemical potential as shown in Table~ \ref{table:table1}.

Fig.~\ref{fig:fig2} shows the $T-\omega$ phase diagram with different chemical potentials $\mu=0, 0.05, 0.08, 0.1 {\rm GeV}$ for G$_V$ = $0$, $0.67$G$_S$ and $-0.5$G$_S$, respectively. The corresponding locations of the CEP are listed in Table~\ref{table:table2}. For the case of G$_V$ = 0 and $\mu=0$, the CEP in the $T-\omega$ plane is located at $T^E=0.035{\rm GeV}, \omega^E=0.663 {\rm GeV}$. Unlike the case in the $T-\mu$ phase diagram, a vector interaction has little effect on changing the phase boundary in the $T-\omega$ plane. The chemical potential does not affect the phase boundary so much in the case of G$_V$ = 0.67G$_S$, however, for the repulsive interaction in the vector channel, the chemical potential has explicit effect on the phase boundary in the $T-\omega$ plane. Similar to the role of the angular velocity on the CEP in the $T-\mu$ plane, the chemical potential only shifts down the critical temperature $T^E$ and does not change the critical angular velocity, which can be read from Table~\ref{table:table2} explicitly.

\begin{table}
\begin{tabular}{|c|c|c|c|}
\hline
~$\{$T$^E$,$\mu^E\}~ $&~ G$_V$=0 ~& ~G$_V$= 0.67 G$_S$~& ~ G$_V$=-0.5 G$_S$ ~ \\
\hline
$\omega = 0$~GeV&\{0.032\text{GeV}, 0.333\text{GeV}\} &--& $\{0.098\text{GeV}, 0.241\text{GeV}\}$  \\
\hline
$\omega = 0.1~$GeV& \{0.002\text{GeV}, 0.333\text{GeV}\} &--&  $\{0.093\text{GeV}, 0.241\text{GeV}\}$   \\
\hline
$\omega = 0.15$~GeV&--& -- &  $\{0.087\text{GeV}, 0.241\text{GeV}\}$  \\
\hline
$\omega = 0.18$ ~GeV&--&--& $\{0.082\text{GeV}, 0.241\text{GeV}\}$     \\
\hline
\end{tabular}
\caption{The locations of CEP in the $T-\mu$ plane with different angular velocities and vector interaction couplings. }
\label{table:table1}
\end{table}

\begin{table}
\begin{tabular}{|c|c|c|c|}
\hline
~$\{$T$^{E}$, $\omega^E\}~ $&~ G$_V$=0 ~& ~G$_V$= 0.67 G$_S$~& ~ G$_V$=-0.5 G$_S$ ~ \\
\hline
$\mu = 0$~GeV&\{0.035\text{GeV}, 0.663\text{GeV}\} & \{0.031\text{GeV}, 0.673\text{GeV}\} &\{0.037\text{GeV}, 0.663\text{GeV}\}\\
\hline
$\mu = 0.05~$GeV&\{0.024\text{GeV}, 0.663\text{GeV}\}& \{0.020\text{GeV}, 0.673\text{GeV}\} & \{0.018\text{GeV}, 0.66\text{GeV}\} \\
\hline
$\mu= 0.08$~GeV&--& \{0.005\text{GeV}, 0.673\text{GeV}\}&-- \\
\hline
$\mu = 0.1$ ~GeV&--&--& -- \\
\hline
\end{tabular}
\caption{The locations of CEP in the $T-\omega$ plane with different chemical potentials and vector interaction couplings.}
\label{table:table2}
\end{table}

\subsection{The effect of vector interaction on the phase structure}

\begin{figure}[t!]
\includegraphics[width=6.5cm]{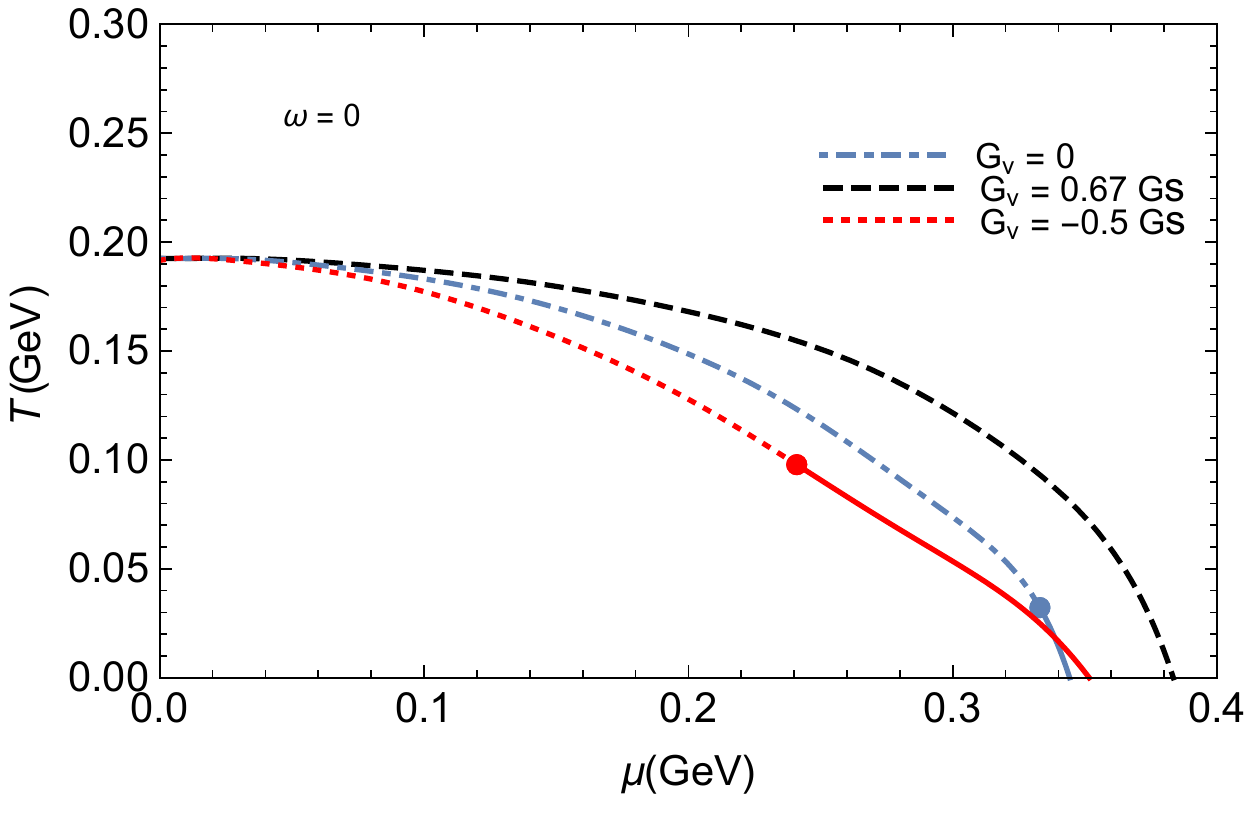}
\includegraphics[width=6.5cm]{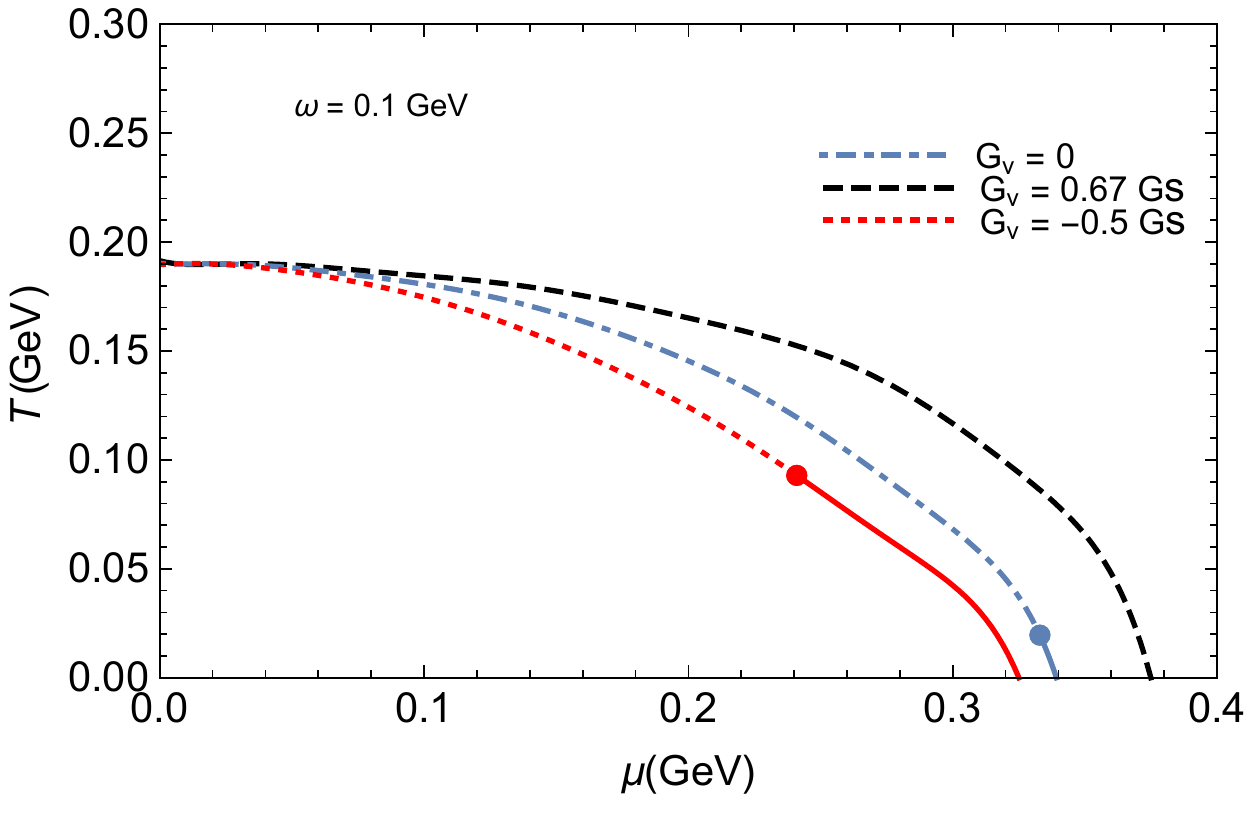}
\includegraphics[width=6.5cm]{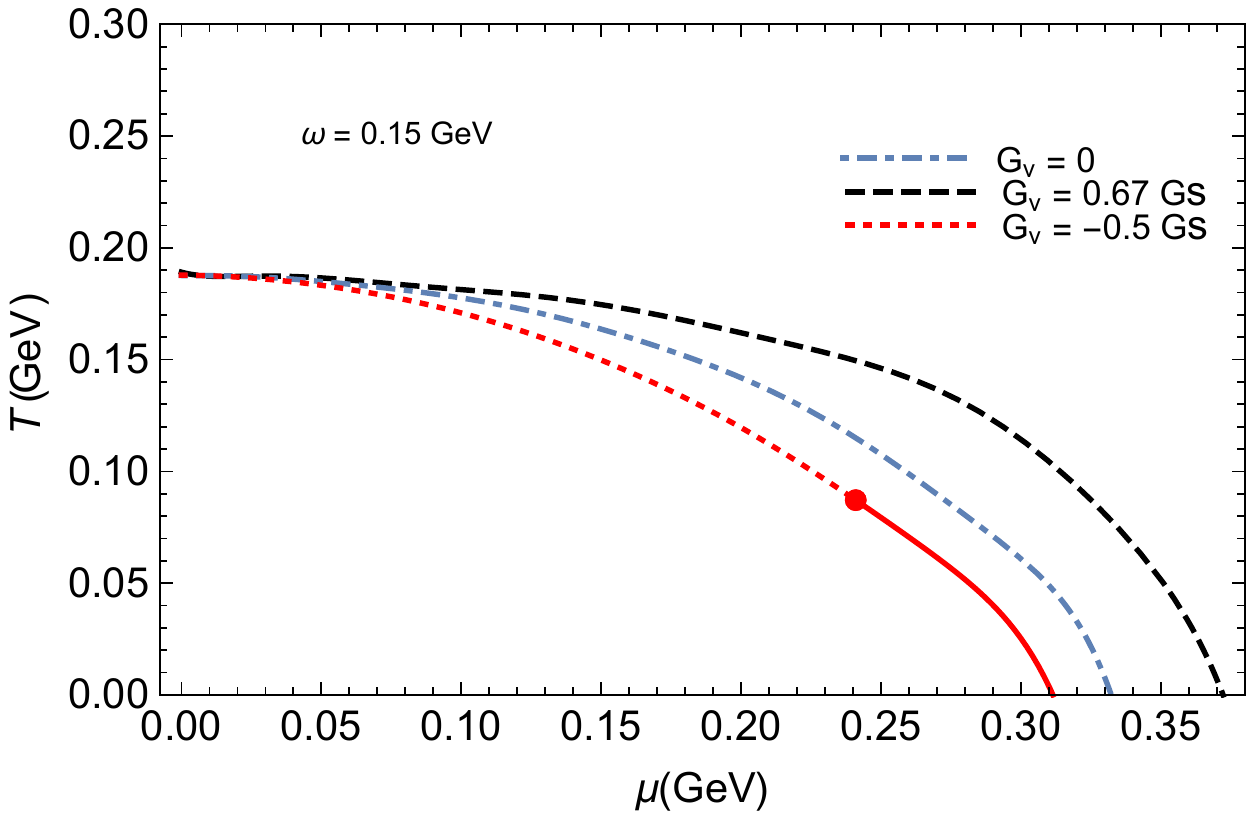}
\includegraphics[width=6.5cm]{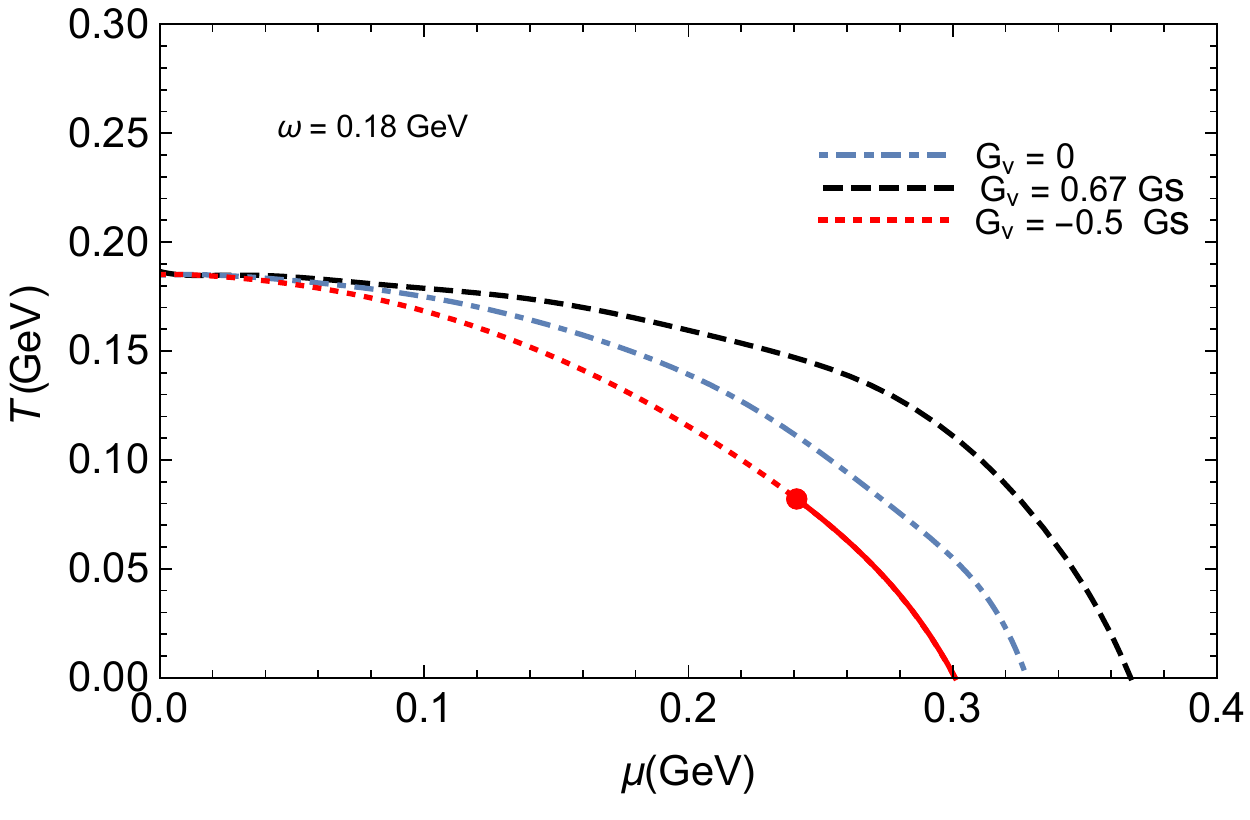}
\caption{The phase diagram in the T-$\mu$ plane with different  G$_V$. The angular velocities are $\omega$ = 0, 0.1 GeV, 0.15 GeV, 0.18 GeV from top left to bottom right.}

\label{fig:fig3}
\end{figure}

\begin{figure}[t!]
\includegraphics[width=6.5cm]{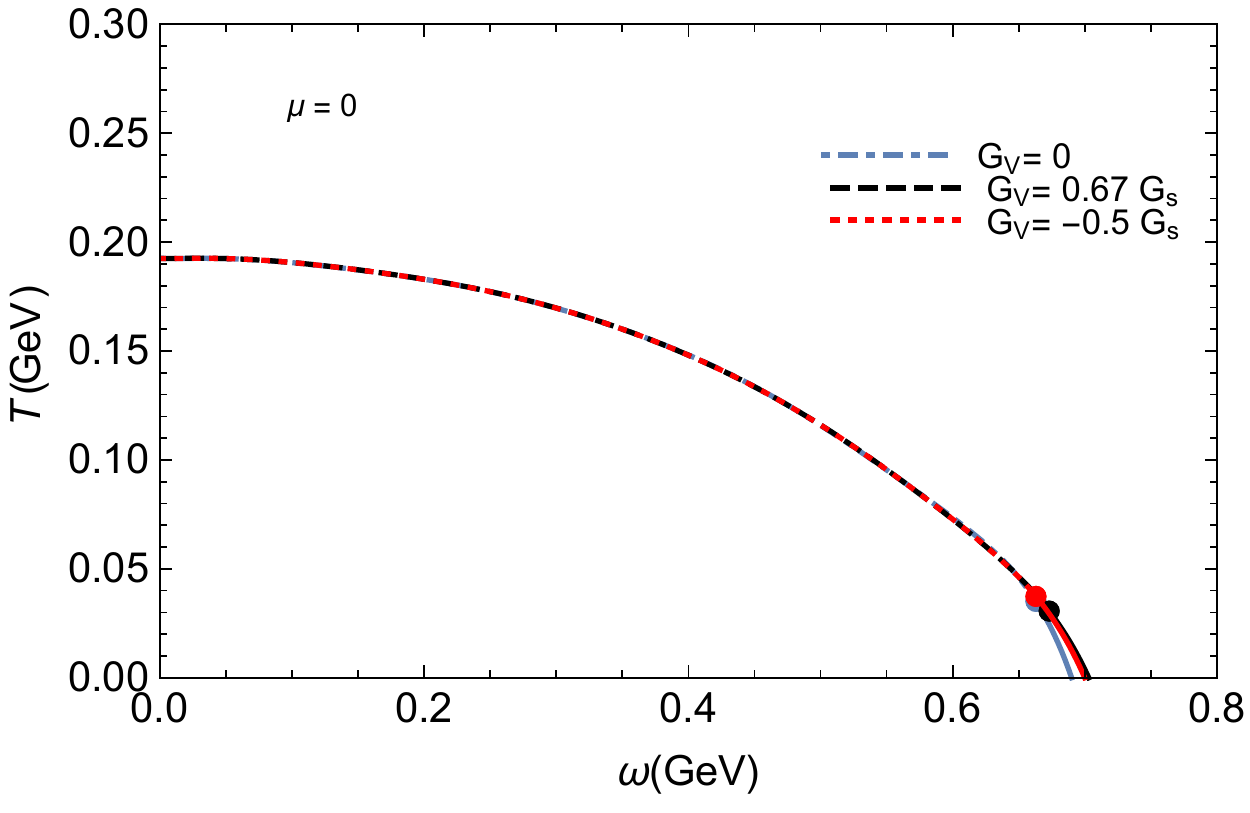}
\includegraphics[width=6.5cm]{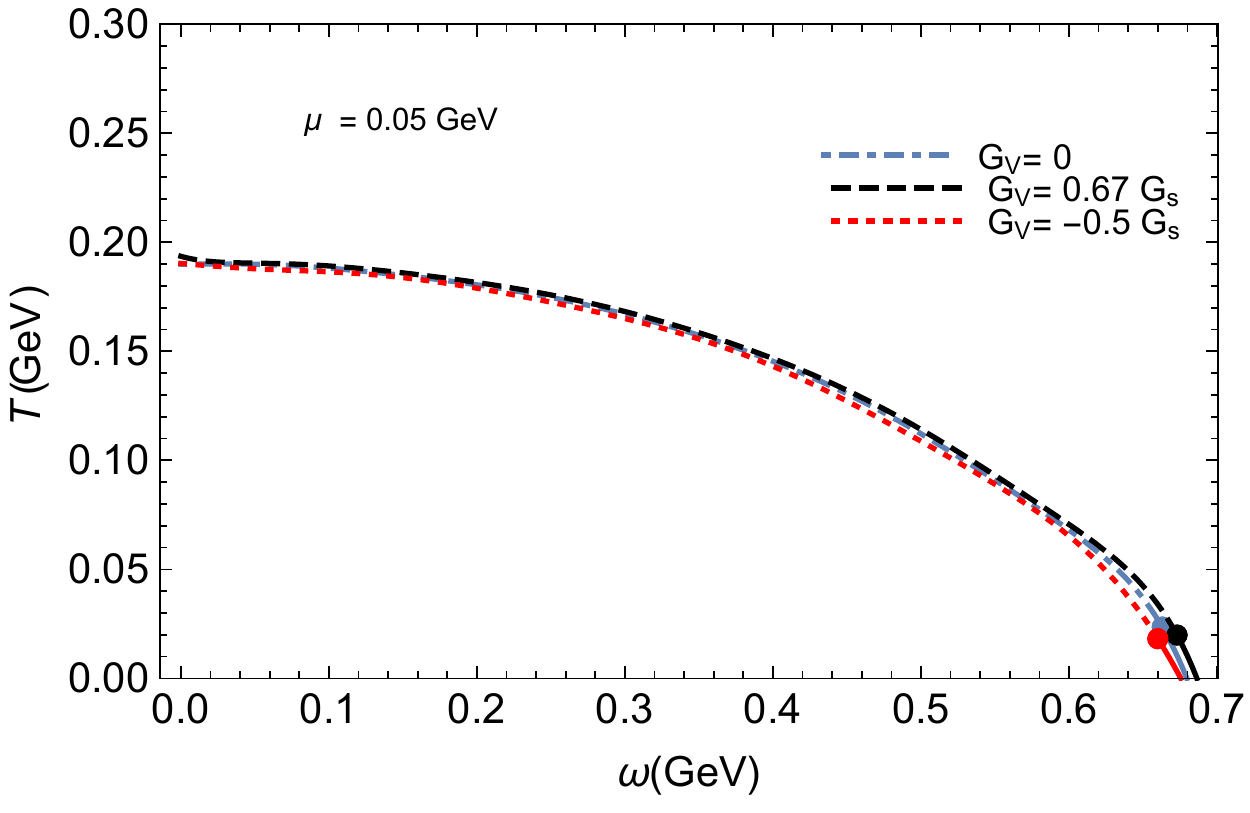}
\includegraphics[width=6.5cm]{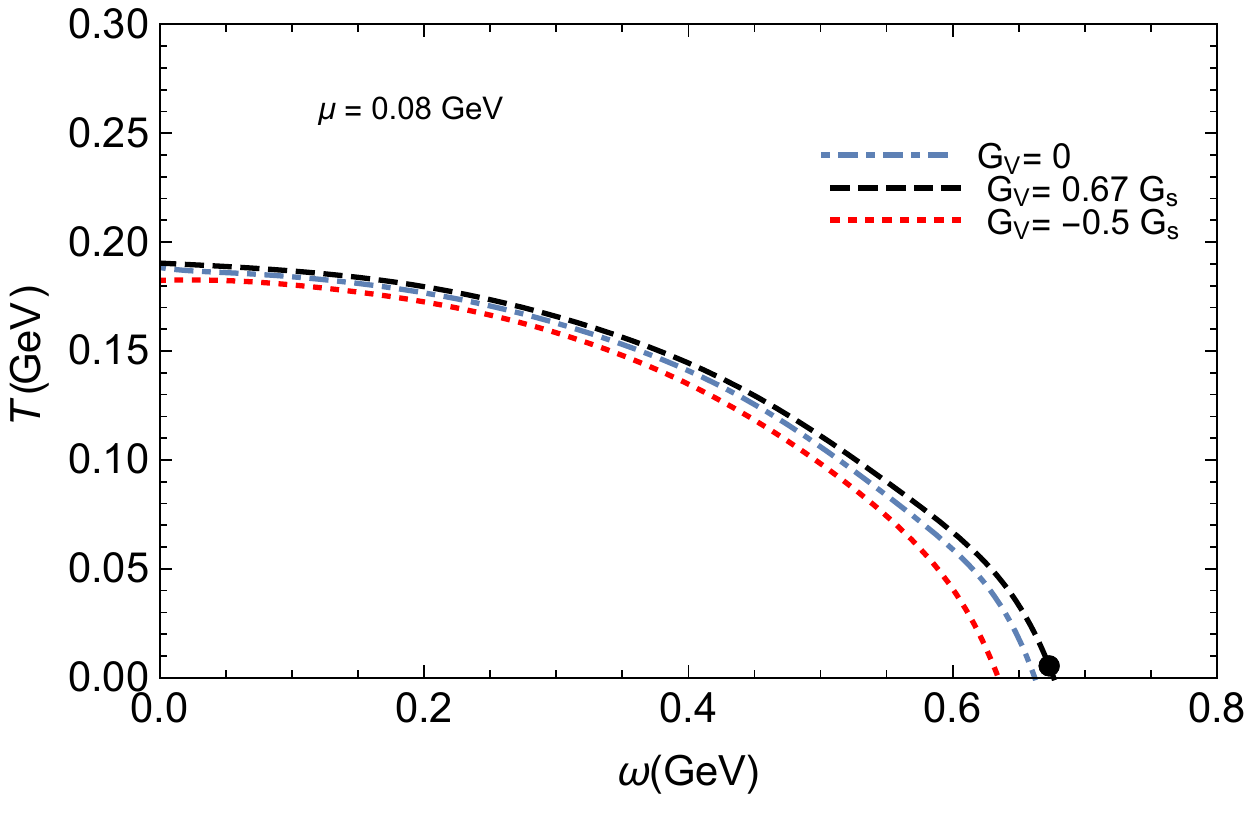}
\includegraphics[width=6.5cm]{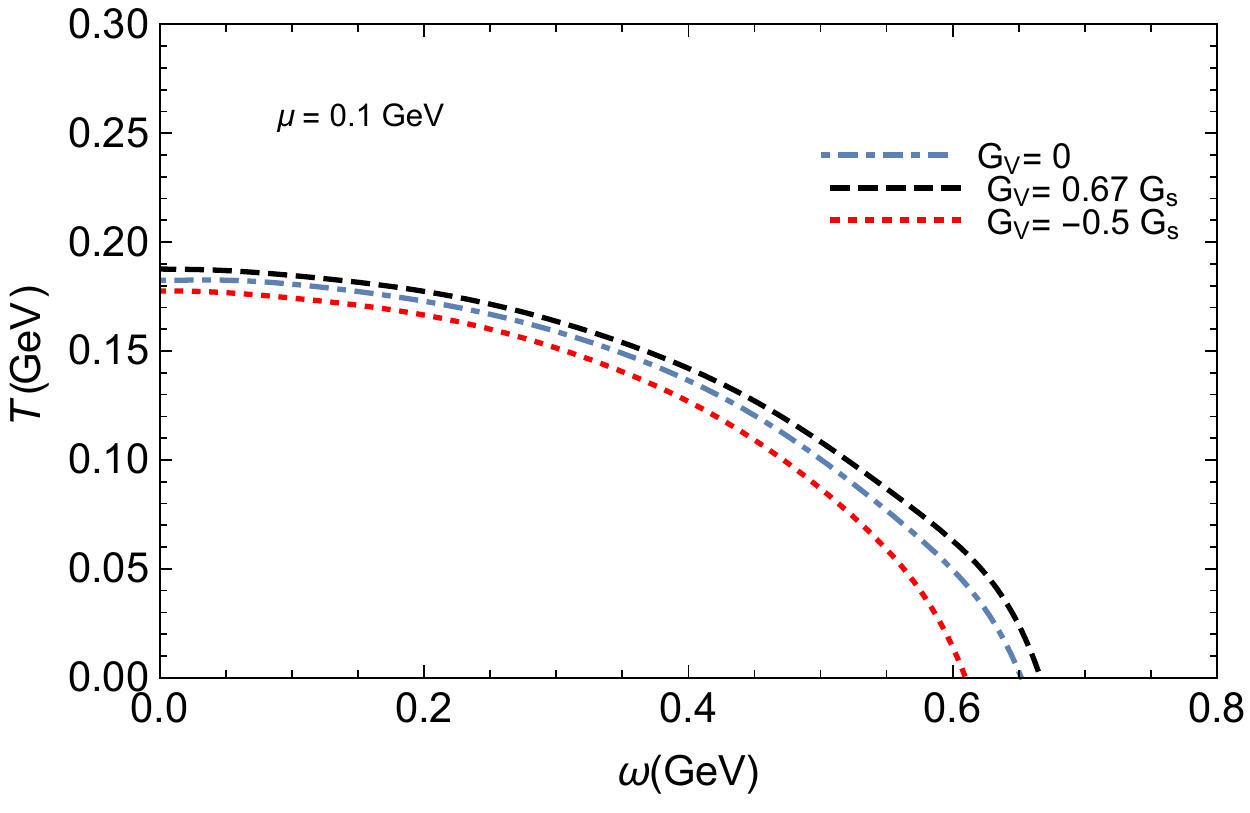}
\caption{The phase diagram in the T-$\omega$ plane with different  G$_V$. The chemical potentials are $\mu$ = 0, 0.05 GeV, 0.08 GeV, 0.1 GeV from top left to bottom. }
\label{fig:fig4}
\end{figure}

In order to see the effect of the vector interaction on the phase structure, we show the phase diagram with different vector interaction G$_V$
in the $T-\mu$ plane and $T-\omega$ plane in Fig.~\ref{fig:fig3} and Fig.~\ref{fig:fig4}, respectively.

Fig.~\ref{fig:fig3} shows the $T-\mu$ phase diagram with different vector interactions G$_V$ = 0, 0.67G$_S$ and -0.5G$_S$, respectively. It is found that
when the coupling constant in the vector channel is positive G$_V$ = 0.67G$_S$, there is no CEP showing up in the $T-\mu$ phase diagram for all the angular velocities  $\omega=0, 0.1  {\rm GeV}, 0.15  {\rm GeV}, 0.18 {\rm GeV}$. In the case of no vector interaction G$_V$ = 0, the CEP shows up in the $T-\mu$ plane for slowly rotating quark matter system, when the system rotates faster, the CEP disappears. If there exists a repulsive interaction in the vector channel, i.e, G$_V$ =-0.5G$_S$, the CEP shows up in the $T-\mu$ phase diagram for all the angular velocities $\omega=0, 0.1  {\rm GeV}, 0.15  {\rm GeV}, 0.18 {\rm GeV}$. The increase of the angular velocity
shifts the CEP to the right part of the phase boundary.

Fig.~\ref{fig:fig4} shows the $T-\omega$ phase diagram with different vector interactions G$_V$ = 0, 0.67G$_S$ and -0.5G$_S$, respectively. For small chemical potentials $\mu=0,0.05 {\rm GeV}$, it is found that the vector interaction almost has no effect on the phase boundary as well as the CEP in the $T-\omega$ plane. When the chemical potential increases to $\mu=0.08 {\rm GeV}$ and $0.1 {\rm GeV}$, the CEP disappears from the $T-\omega$ plane, and the vector interaction has larger effect on the phase diagram with the increase of the chemical potential.


\subsection{The 3D phase diagram in the $T-\mu-\omega$ plane}

\begin{figure}[t!]
\centerline{\includegraphics[width=8cm]{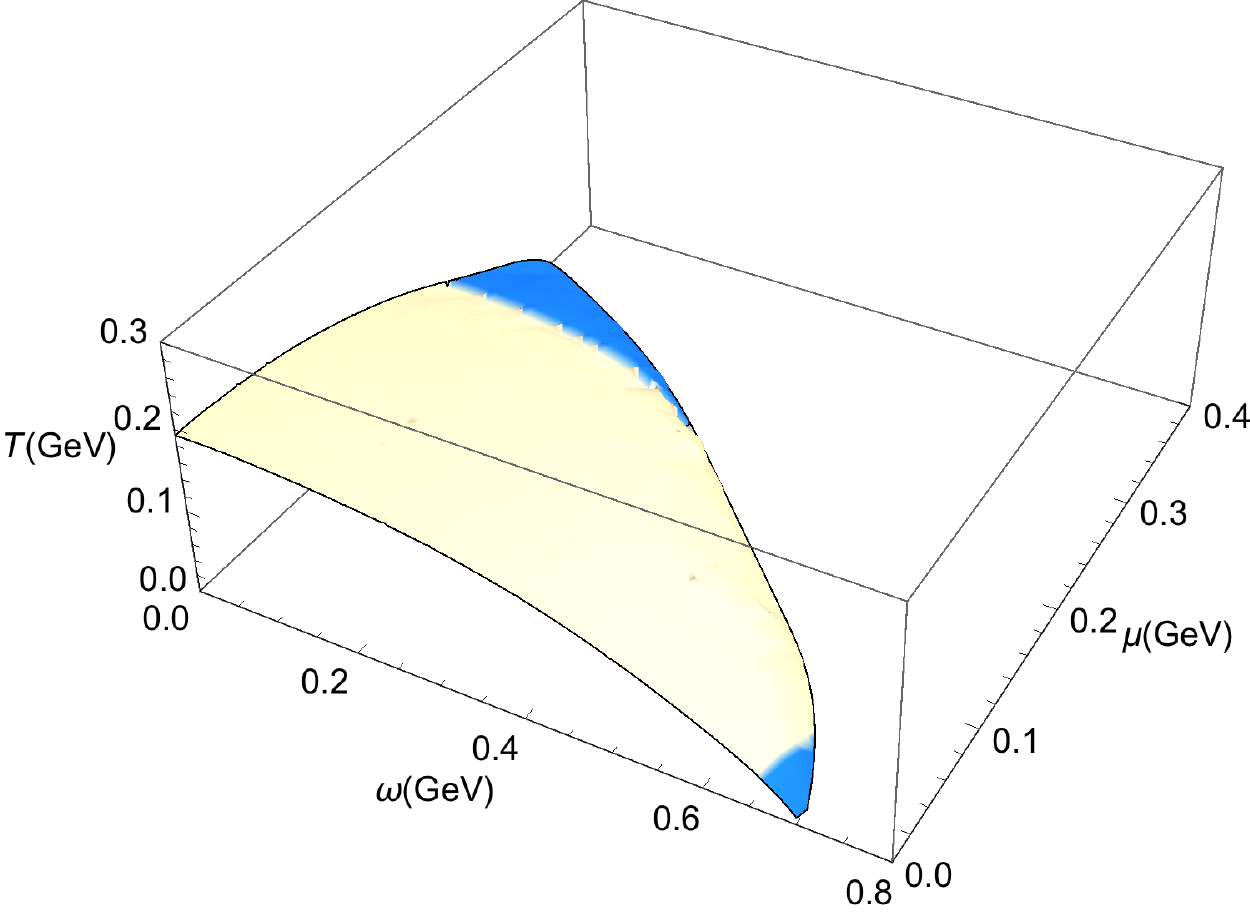}}
\caption{The 3D phase structure for chiral transition on $(T,\mu,\omega)$ frame with G$_V$ = -0.5G$_S$.}
\label{fig:fig5}
\end{figure}

We show the 3D phase diagram in the $T-\mu-\omega$ plane in Fig.~\ref{fig:fig5} with the coupling constant in the vector channel G$_V$=-0.5 G$_S$.
Because the properties we have observed that the angular velocity only shifts down the critical temperature of CEP in the $T-\mu$ plane,
and the chemical potential only shifts down the critical temperature of CEP in the $T-\omega$ plane, we can explicitly see that most part
on the phase diagram is crossover, and the first order chiral phase transition only exists in two corners on the surface, i.e. in the corner of small $\omega$ and large $\mu$ and the corner of small $\mu$ and large $\omega$, as shown by blue regions on the graph. There are two obvious boundaries at around $\omega \approx 0.66 {\rm GeV}$ and $\mu \approx 0.24 {\rm GeV}$.

\begin{figure}[t!]
\centerline{\includegraphics[width=8cm]{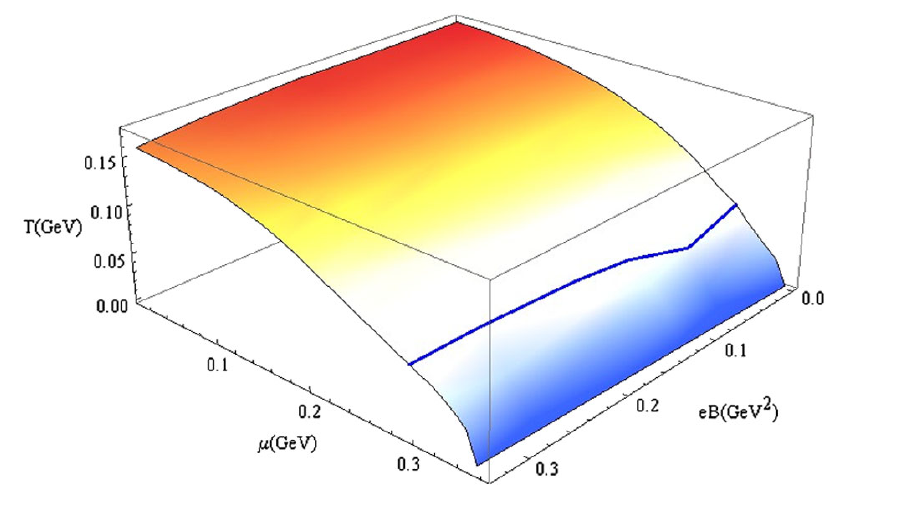}}
\caption{The 3D phase structure for chiral transition in the $(T,\mu,eB)$ frame taken from~\cite{Chao:2013qpa}.}
\label{fig:fig6}
\end{figure}

By comparing with the 3D phase structure for chiral phase transition in $(T,\mu,eB)$ frame as shown in Fig.~\ref{fig:fig6} taken from~\cite{Chao:2013qpa}, we can see that the angular momentum plays a quite different role as the magnetic field.  From previous studies we know the phase diagram in a external magnetic field which is influenced  by two main mechanisms: the magnetic catalysis which enhances the chiral symmetry breaking in the vacuum and the inverse magnetic catalysis which helps chiral symmetry restoration around the critical temperature. The solid line in Fig.~\ref{fig:fig6} is for the critical end point (CEP).
taken into account the inverse magnetic catalysis effect, the location of CEP $(T^E, \mu^E)$ does not change so much at high magnetic fields, which is
different from the result in~\cite{Menezes:2009uc} with only magnetic catalysis and the location of the CEP moves towards the temperature-axis with increasing magnetic field. In the case of angular momentum, it only helps the chiral symmetry restoration. In fact in Eq.~(\ref{omg}), the main contribution of the thermodynamic potential is given by $n=0$ term.  It is observed that approximately the angular velocity only gives an addition contribution to the dynamical chemical potential $\tilde{\mu}$, which means that the angular velocity and chemical potential are approximately equivalent.

\subsection{The baryon number susceptibilities}

The baryon number fluctuations are sensitive to the CEP, which has been discussed in many literatures, for review, see \cite{Luo:2017faz}. The cumulants of conserved quantities up to fourth order of net-proton have been measured in the first phase of beam energy scan program (BES-I) at RHIC for Au+Au collisions
\cite{Adamczyk:2013dal,Aggarwal:2010wy,Luo:2017faz}, and a non-monotonic energy dependent behavior for the kurtosis of the net proton
number distributions $\kappa \sigma^2$ has been observed. As we mentioned in the introduction, that the created matter at relativistic heavy ion collisions is also fast rotating, therefore it is very important to estimate how large the rotation will affect on the baryon number fluctuations.

In this part, we investigate the effect of the angular velocity on the kurtosis of the baryon number fluctuation $\kappa\sigma^2$, which is defined as
\eqn
\kappa \sigma^2 =\frac{\chi_4^B}{\chi_2^B},
\eeqn
with
\eqn
\chi_n^B = \frac{\partial^n(P/T^4)}{\partial(\mu_B/T)^n},
\eeqn
where the pressure P = $-\Omega$ is just the minus of the the grand potential.

\begin{figure}[t!]
\begin{center}
\includegraphics[width=6.5cm]{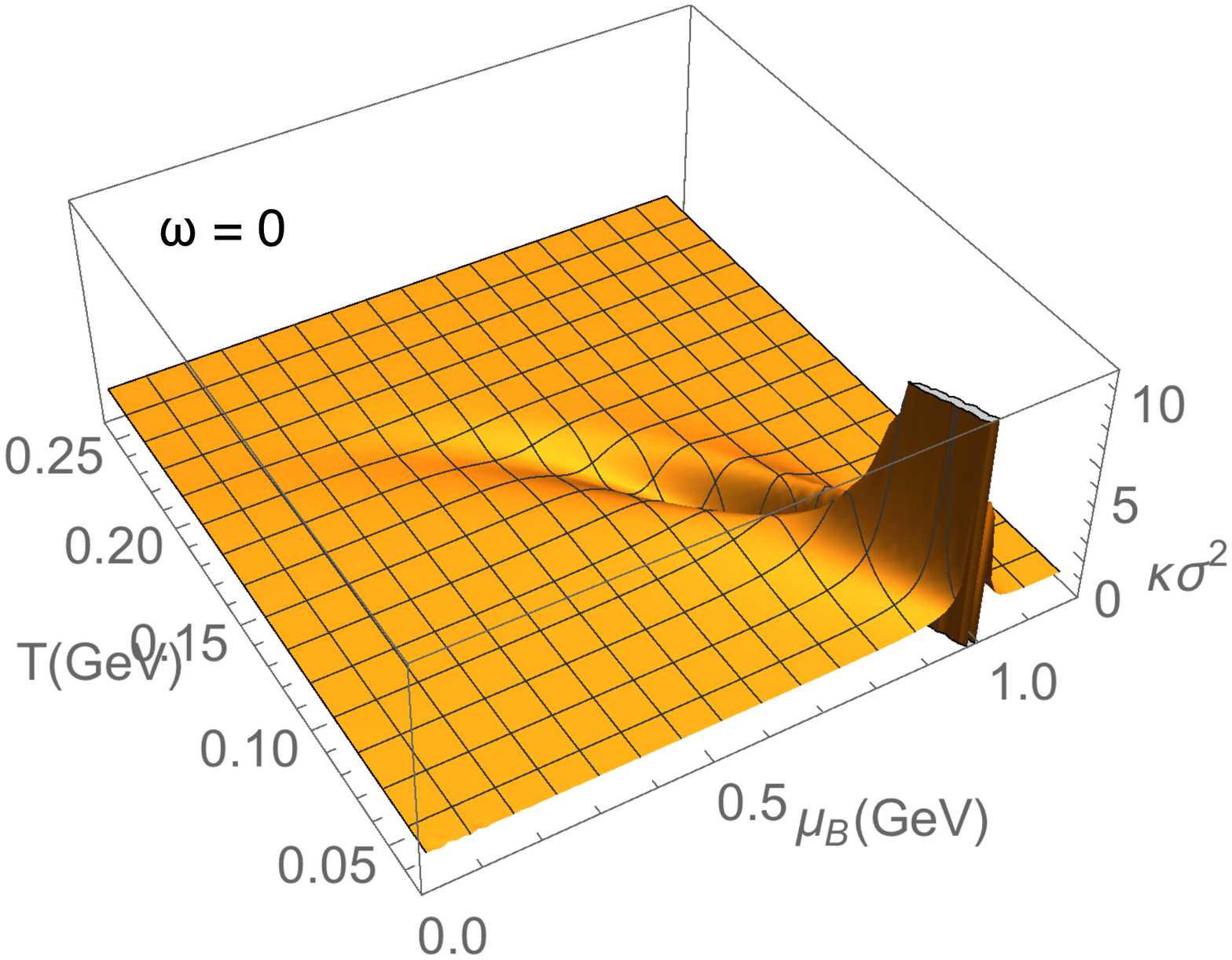}
\includegraphics[width=6.5cm]{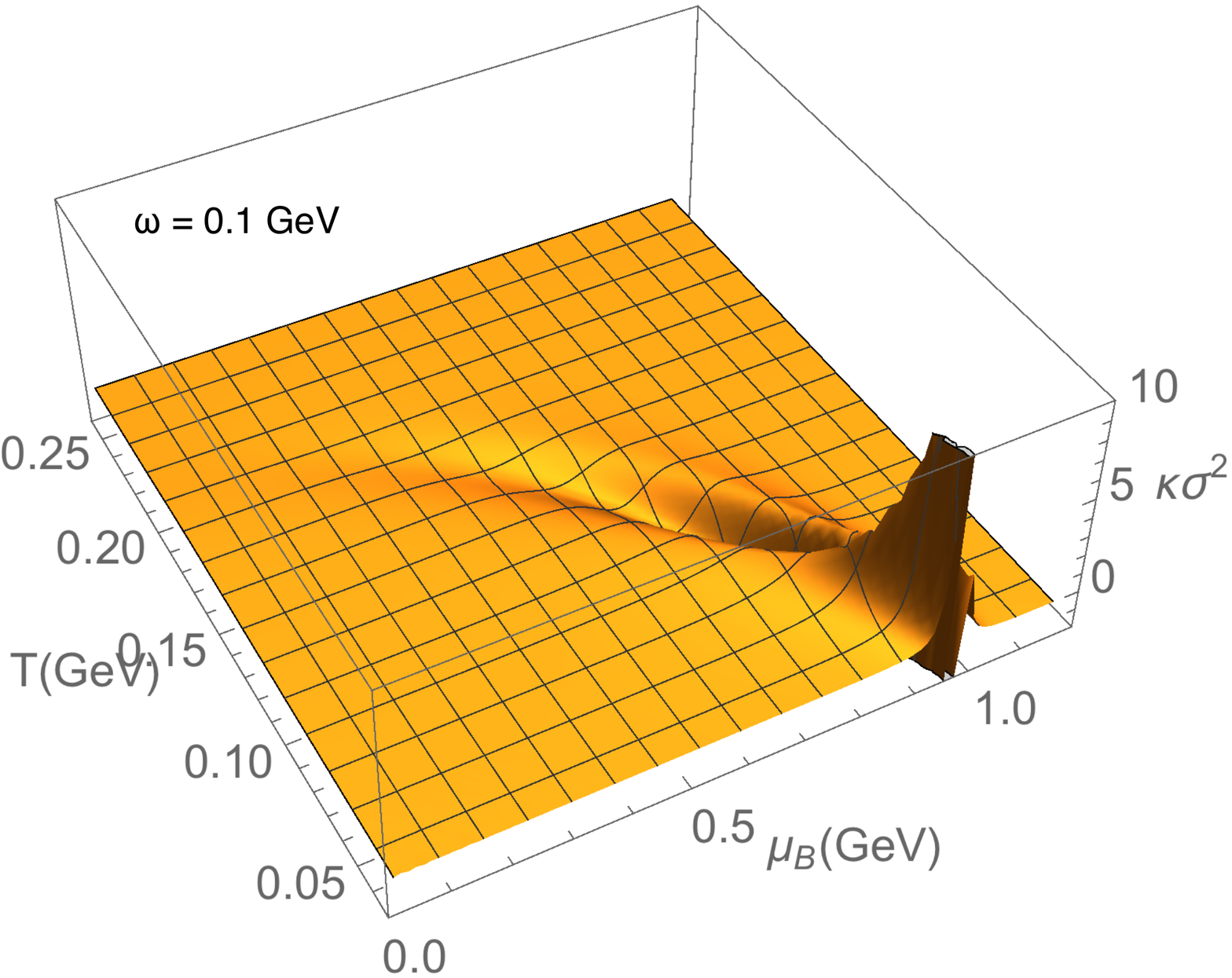}
\vskip -3cm
\includegraphics[width=6.5cm]{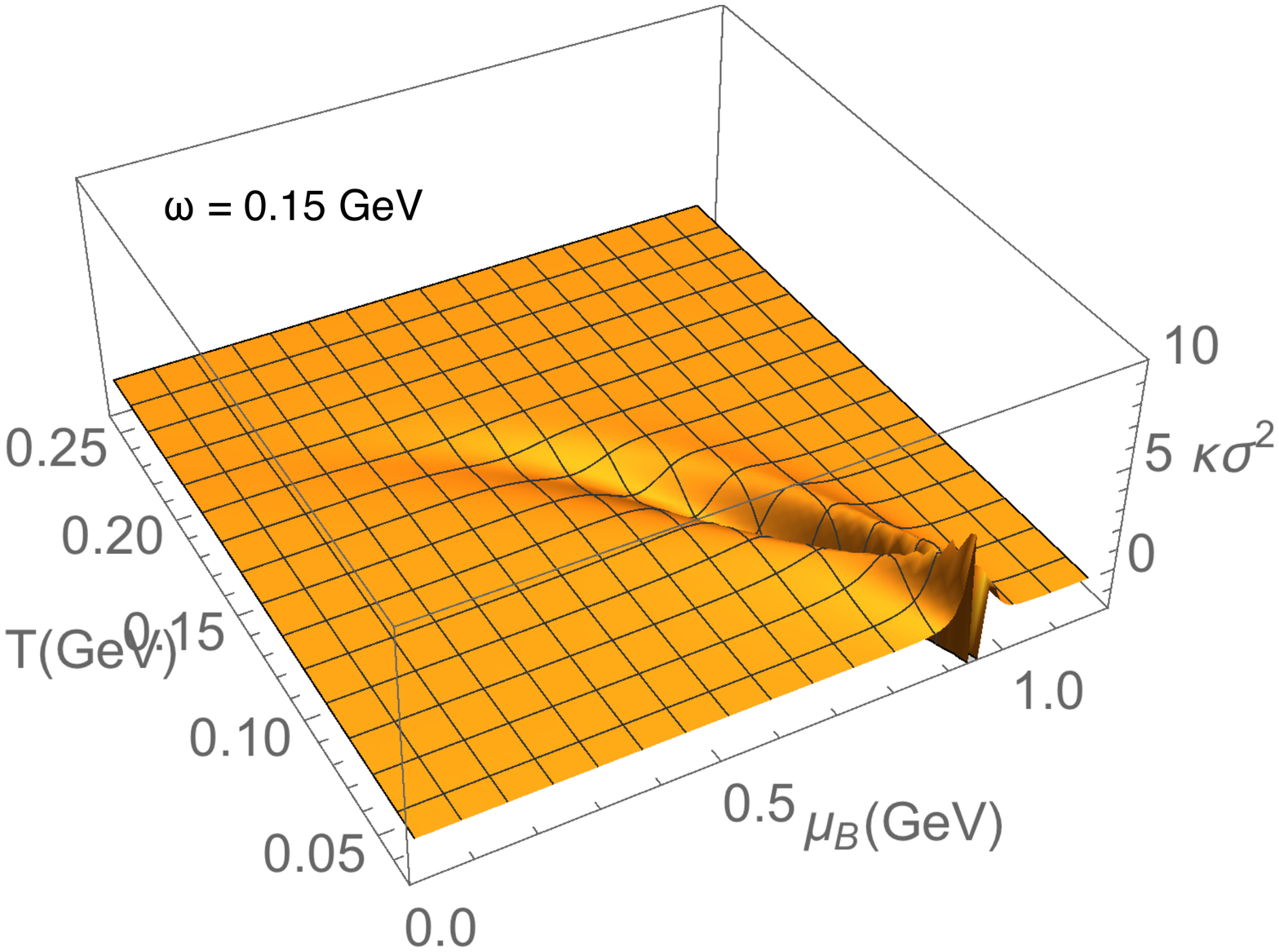}
\includegraphics[width=6.5cm]{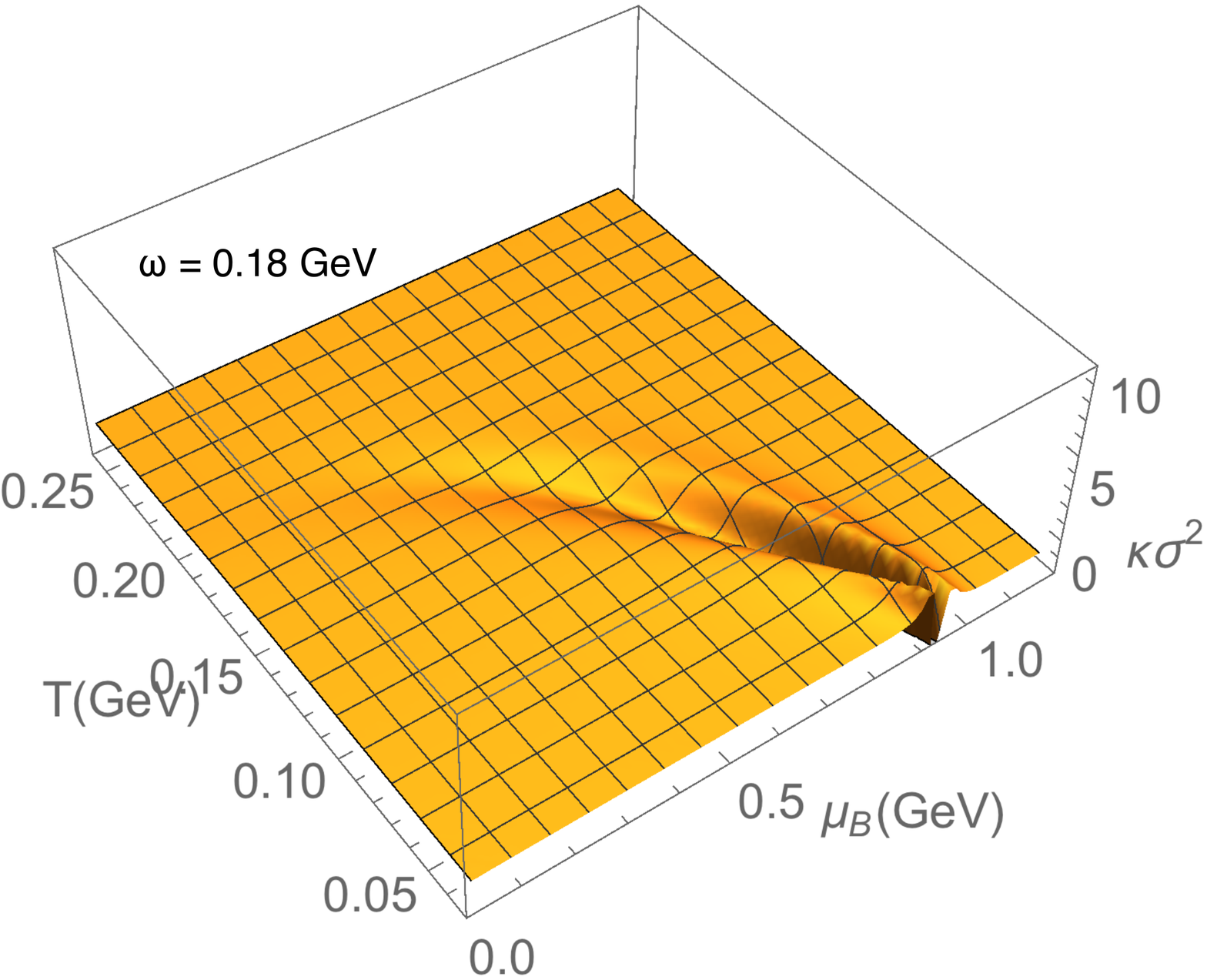}
\caption{The kurtosis of baryon number fluctuation $\kappa\sigma^2$ with different angular velocity $\omega = 0, 0.1  {\rm GeV}, 0.15 {\rm GeV} , 0.18 {\rm GeV}$.}
\label{fig:fig7}
\end{center}
\end{figure}

\begin{figure}[t!]
\begin{center}
\includegraphics[width=8.5cm]{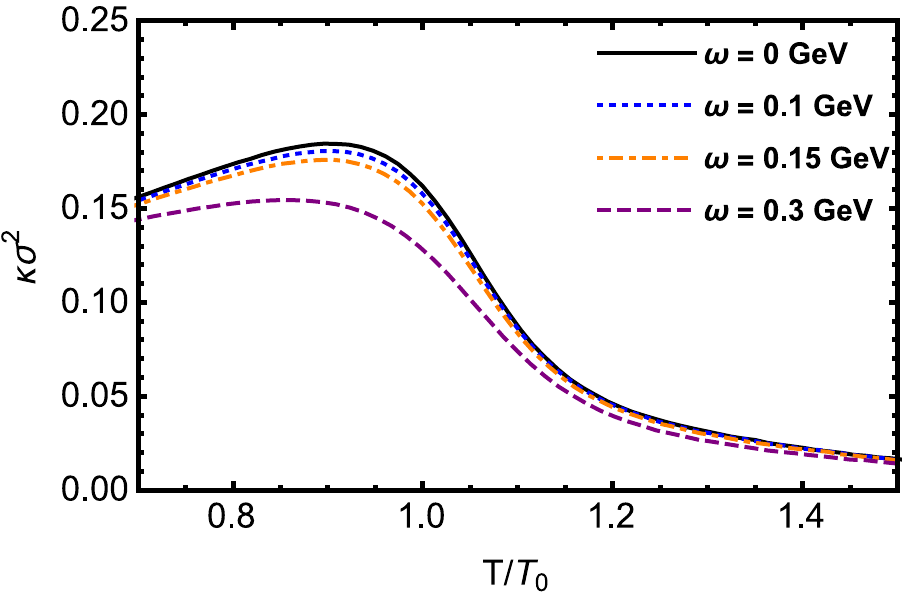}
\caption{The kurtosis of baryon number fluctuation $\kappa\sigma^2$ with different angular velocity $\omega = 0, 0.1  {\rm GeV}, 0.15 {\rm GeV} , 0.3 {\rm GeV}$.}
\label{fig:fig8}
\end{center}
\end{figure}

In Fig.~\ref{fig:fig7} we show the 3D plot for the kurtosis of baryon number fluctuation $\kappa \sigma^2$ as a function of the temperature and baryon chemical potential with different angular velocities in the NJL model in the case of  G$_ V=0$. We can clearly see that the angular velocity shifts the
location of the CEP to the right part of the phase diagram in the $(T,\mu)$ plane, and with the further increase of the angular velocity, the CEP disappears.
In order to show how the angular velocity affects the baryon number fluctuation, we show in Fig.~\ref{fig:fig8} the value of $\kappa \sigma^2$ as a function of normalized temperature $T/T_0$ in different angular velocity at zero chemical potential, with $T_0$ the critical temperature for chiral phase transition at $\mu=0$. As we have discussed in Ref.~\cite{Li:2018ygx} that in the case of comparing with lattice result Ref.~\cite{Bazavov:2017dus}, quark dynamics only contributes around $20\%$ to the $\kappa \sigma^2$ and the main contribution comes from the gluodynamics, therefore the value of $\kappa \sigma^2$  is only 0.15 at chiral phase transition $T/T_0=1$ in the NJL model. We can see explicitly that the angular velocity decreases the kurtosis of the baryon number fluctuation $\kappa \sigma^2$.

\subsection{Other thermodynamics quantities}

To complete our study on the effect of rotation on QCD matter, in this part, we investigate the equation of state of the rotating matter.
The energy density $\epsilon$ is given by,
\eqn
\epsilon = -T^2\left. \frac{\partial(\Omega/T)}{\partial T}\right|_{V}=-T\left.\frac{\partial \Omega}{\partial T}\right|_{V}+\Omega,
\eeqn
and the corresponding specific heat
\eqn
C_{V} =\left.\frac{\partial \epsilon}{\partial T}\right|_{V} = -T\left.\frac{\partial^2 \Omega}{\partial T^2}\right|_{V}.
\eeqn
The square of velocity of sound at constant entropy S is given by
\eqn
v_s^2 = \left.\frac{\partial P}{\partial \epsilon}\right|_S = \left.\frac{\partial P}{\partial T}\right|_V\bigg{/}\left.\frac{\partial \epsilon}{\partial T}\right|_V = \left.\frac{\partial \Omega}{\partial T}\right|_V\bigg{/}\left.T\frac{\partial^2 \Omega}{\partial T^2}\right|_V.
\eeqn

\begin{figure}[t!]
\begin{center}
\includegraphics[width=6.5cm]{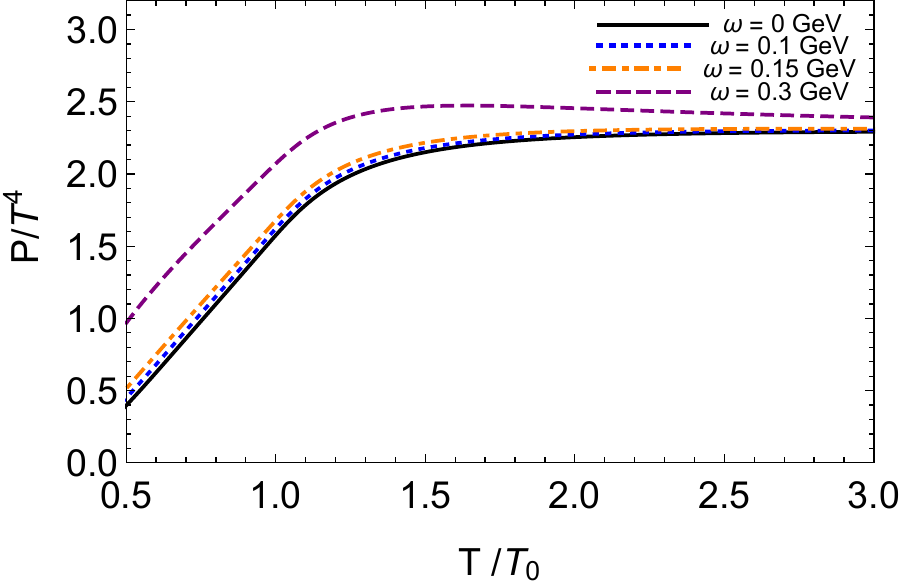}
\includegraphics[width=6.5cm]{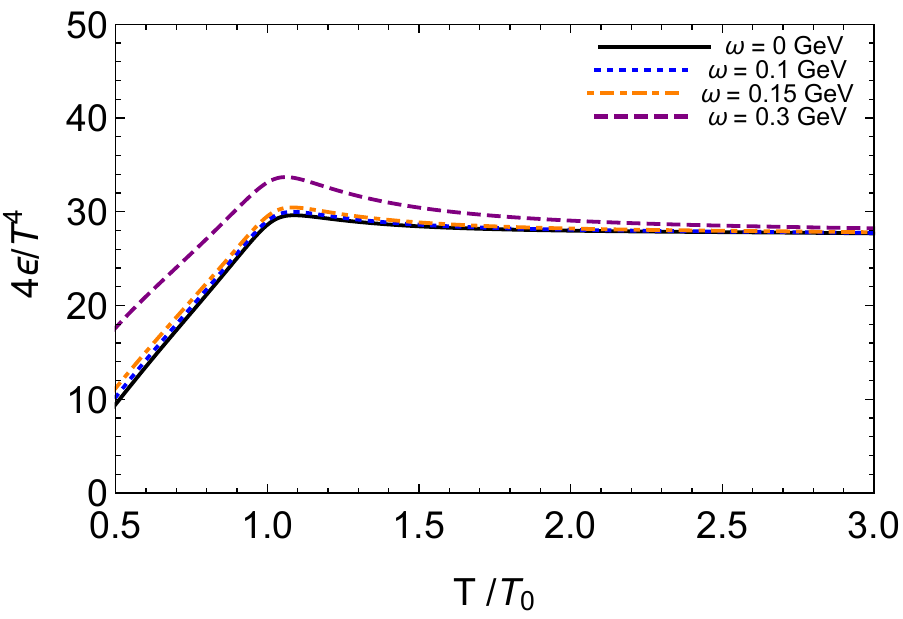}
\includegraphics[width=6.5cm]{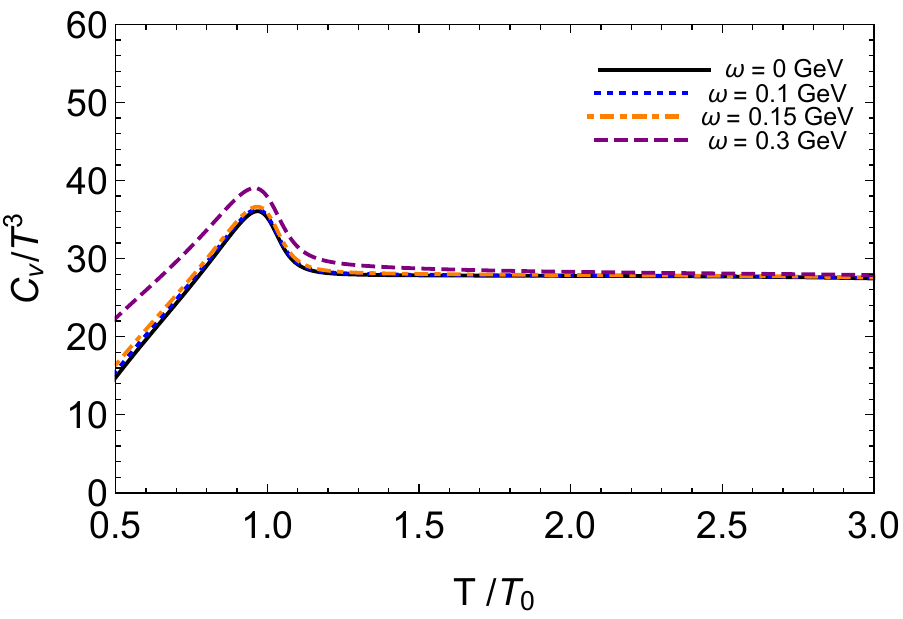}
\includegraphics[width=6.5cm]{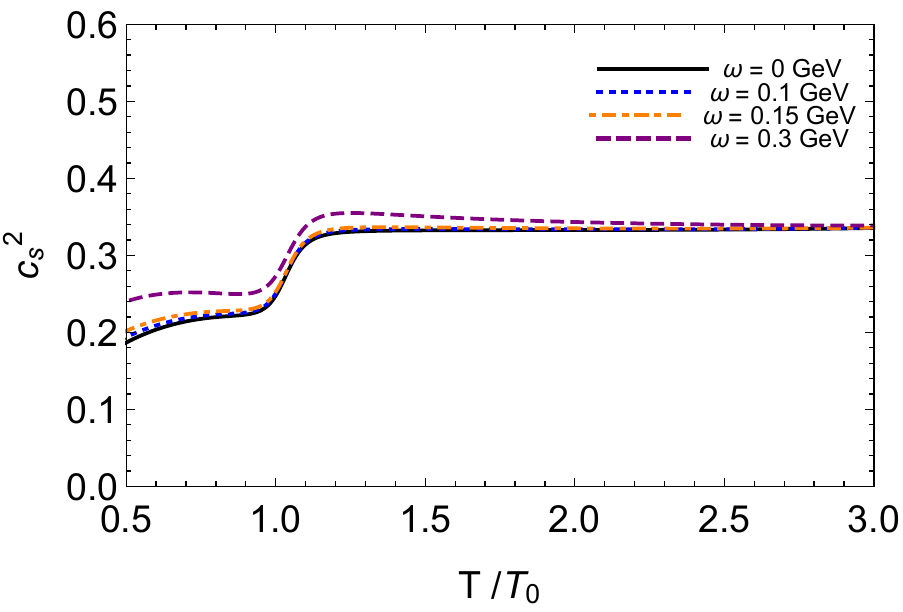}
\caption{$P$, $\epsilon$, $C_{V}$ and $v_s^2$ with different angular velocity $\omega = 0, 0.1  {\rm GeV}, 0.15 {\rm GeV} , 0.3 {\rm GeV}$ at zero chemical potential.}
\label{fig:fig9}
\end{center}
\end{figure}

In Fig.~\ref{fig:fig9} we show the dimensionless quantities of the pressure, energy density, specific heat and the square of sound velocity as a function of normalized temperature $T/T_0$ with different angular velocities at zero chemical potential and zero vector interaction in the NJL model. When the normalized temperature is small, it clearly shows that the larger the angular velocity is, the larger these quantities will be. At high temperatures, the angular velocity plays almost no effect on the rotating matter, and all these quantities reach to the free ideal gas limit.

\section{Conclusion and Outlook}
\label{Conclusion}

In this work, we investigate the effect of the angular velocity on the chiral phase transition of quark matter in the two-flavor NJL model with vector interaction.
It is found that the angular momentum plays similar role as the baryon chemical potential, which suppresses the chiral condensate and helps the chiral phase transition. Therefore, the chiral phase transition shows a critical end point not only in the temperature-chemical potential $T-\mu$ plane, but also in the temperature-angular momentum $T-\omega$ plane. One interesting observation is that in the $T-\mu$ plane, the presence of the angular momentum only shifts down the critical temperature $T^E$ of the CEP and does not shift the critical chemical potential $\mu^E$, and in the $T-\omega$ plane, the increase of the chemical potential only shift down the critical temperature $T^E$ and does not change the critical angular momentum $\omega^E$. From the 3D phase structure in the $(T,\mu,\omega)$ frame, we can explicitly see that most part on the phase diagram is crossover, and the first order chiral phase transition only exists in two corners on the surface, i.e. in the corner of small $\omega$ and large $\mu$ and the corner of small $\mu$ and large $\omega$. By comparing with the 3D phase structure for chiral phase transition in $(T,\mu,eB)$ frame, we can see that the angular momentum plays a quite different role comparing with the magnetic field, because the magnetic field enhances the chiral symmetry breaking in the vacuum which is called the magnetic catalysis effect, and helps chiral symmetry restoration around the critical temperature which is called the inverse magnetic catalysis effect.

The effect of the vector interaction is also investigated, and it is found that the phase structure in $T-\mu$ plane is sensitive to the coupling strength in the vector channel, while the phase structure in $T-\omega$ plane is not sensitive to the vector interaction. The baryon number fluctuations is also investigated and it is found that the angular velocity suppresses the kurtosis of the baryon number fluctuation. In fact, we could see that the critical angular velocity in the $T-\omega$ plane is very high ($\sim0.7 {\rm GeV}$) comparing with the angular velocity which could be created in the heavy ion collision ($\sim 0.1 {\rm GeV}$). So it is hard to find the critical end point in the $T-\Omega$ plane through heavy ion collision experiments, but the influence of the angular velocity on the phase boundary in the $T-\mu$ plane is still obvious, especially in the high baryon density region. This also affects the kurtosis of the baryon number fluctuation in the heavy ion collision. It is also observed that the rotating angular velocity enhances the pressure density, energy density, the specific heat and the sound velocity.

The NJL model only captures quark dynamics, therefore in this work we only consider the properties of quark matter under rotation. In the future, it is worth to study the properties of gluonic matter under fast rotation.

\begin{acknowledgments}
We thank Hui Liu and Lang Yu for useful discussion. The work is supported by the NSFC under Grant Nos. 11725523, 11735007 and 11261130311(CRC 110 by DFG and NSFC), Chinese Academy of Sciences under Grant No. XDPB09, and the start-up funding from UCAS.
\end{acknowledgments}


\begin{thebibliography}{99}

\bibitem{Fukushima:2008xe}
  K.~Fukushima, D.~E.~Kharzeev and H.~J.~Warringa,
  Phys.\ Rev.\  D {\bf 78}, 074033 (2008).

\bibitem{Kharzeev:2007tn}
  D.~Kharzeev and A.~Zhitnitsky,
  Nucl.\ Phys.\ A {\bf 797}, 67 (2007).

\bibitem{Kharzeev:2007jp}
  D.~E.~Kharzeev, L.~D.~McLerran and H.~J.~Warringa,
  Nucl.\ Phys.\  A {\bf 803}, 227 (2008).

\bibitem{Klevansky:1989vi}
  S.~P.~Klevansky and R.~H.~Lemmer,
  Phys.\ Rev.\ D {\bf 39}, 3478 (1989).
  doi:10.1103/PhysRevD.39.3478

\bibitem{Klimenko:1990rh}
  K.~G.~Klimenko,
  Theor.\ Math.\ Phys.\  {\bf 89}, 1161 (1992)
  [Teor.\ Mat.\ Fiz.\  {\bf 89}, 211 (1991)].
  doi:10.1007/BF01015908

\bibitem{Gusynin:1995nb}
  V.~P.~Gusynin, V.~A.~Miransky and I.~A.~Shovkovy,
  Nucl.\ Phys.\ B {\bf 462}, 249 (1996)
  doi:10.1016/0550-3213(96)00021-1
  [hep-ph/9509320];
  V.~P.~Gusynin, V.~A.~Miransky and I.~A.~Shovkovy,
  Nucl.\ Phys.\ B {\bf 563}, 361 (1999)
  doi:10.1016/S0550-3213(99)00573-8
  [hep-ph/9908320].

\bibitem{Bali:20111213}
  G.~S.~Bali, F.~Bruckmann, G.~Endrodi, Z.~Fodor, S.~D.~Katz, S.~Krieg, A.~Schafer and K.~K.~Szabo,
  JHEP {\bf 1202}, 044 (2012);
  G.~S.~Bali, F.~Bruckmann, G.~Endrodi, Z.~Fodor, S.~D.~Katz and A.~Schafer,
  Phys.\ Rev.\ D {\bf 86}, 071502 (2012);
 G.~S.~Bali, F.~Bruckmann, G.~Endrodi, F.~Gruber and A.~Schaefer,
  JHEP {\bf 1304}, 130 (2013).
\bibitem{Chernodub:2010qx}
  M.~N.~Chernodub,
  Phys.\ Rev.\ D {\bf 82}, 085011 (2010)
  doi:10.1103/PhysRevD.82.085011
  [arXiv:1008.1055 [hep-ph]].

\bibitem{Chernodub:2011mc}
  M.~N.~Chernodub,
  Phys.\ Rev.\ Lett.\  {\bf 106}, 142003 (2011)
  doi:10.1103/PhysRevLett.106.142003
  [arXiv:1101.0117 [hep-ph]].

\bibitem{Jiang:2016woz}
  Y.~Jiang, Z.~W.~Lin and J.~Liao,
  Phys.\ Rev.\ C {\bf 94}, no. 4, 044910 (2016)
  Erratum: [Phys.\ Rev.\ C {\bf 95}, no. 4, 049904 (2017)]
  doi:10.1103/PhysRevC.94.044910, 10.1103/PhysRevC.95.049904
  [arXiv:1602.06580 [hep-ph]].
\bibitem{Deng:2016gyh}
  W.~T.~Deng and X.~G.~Huang,
  Phys.\ Rev.\ C {\bf 93}, no. 6, 064907 (2016)
  doi:10.1103/PhysRevC.93.064907
  [arXiv:1603.06117 [nucl-th]].
  %
\bibitem{Son:2009tf}
  D.~T.~Son and P.~Surowka,
  Phys.\ Rev.\ Lett.\  {\bf 103}, 191601 (2009).

\bibitem{Kharzeev:2010gr}
  D.~E.~Kharzeev and D.~T.~Son,
  Phys.\ Rev.\ Lett.\  {\bf 106}, 062301 (2011).

\bibitem{Jiang:2015cva}
  Y.~Jiang, X.~G.~Huang and J.~Liao,
  Phys.\ Rev.\ D {\bf 92}, no. 7, 071501 (2015).

\bibitem{Kharzeev:2010gd}
  D.~E.~Kharzeev and H.~U.~Yee,
  Phys.\ Rev.\ D {\bf 83}, 085007 (2011)
  doi:10.1103/PhysRevD.83.085007
  [arXiv:1012.6026 [hep-th]].

  \bibitem{Burnier:2011bf}
  Y.~Burnier, D.~E.~Kharzeev, J.~Liao and H.~U.~Yee,
  Phys.\ Rev.\ Lett.\  {\bf 107}, 052303 (2011); arXiv:1208.2537 [hep-ph].
\bibitem{Andersen:2014xxa}
  J.~O.~Andersen, W.~R.~Naylor and A.~Tranberg,
  Rev.\ Mod.\ Phys.\  {\bf 88}, 025001 (2016)
  doi:10.1103/RevModPhys.88.025001
  [arXiv:1411.7176 [hep-ph]].


\bibitem{Chen:2015hfc}
  H.~L.~Chen, K.~Fukushima, X.~G.~Huang and K.~Mameda,
  Phys.\ Rev.\ D {\bf 93}, no. 10, 104052 (2016)
  doi:10.1103/PhysRevD.93.104052
  [arXiv:1512.08974 [hep-ph]].
  \bibitem{Jiang:2016wvv}
  Y.~Jiang and J.~Liao,
  Phys.\ Rev.\ Lett.\  {\bf 117}, no. 19, 192302 (2016)
  [arXiv:1606.03808 [hep-ph]].

\bibitem{Ebihara:2016fwa}
  S.~Ebihara, K.~Fukushima and K.~Mameda,
  Phys.\ Lett.\ B {\bf 764}, 94 (2017)
  doi:10.1016/j.physletb.2016.11.010
  [arXiv:1608.00336 [hep-ph]].

\bibitem{Chernodub:2016kxh}
  M.~N.~Chernodub and S.~Gongyo,
  JHEP {\bf 1701}, 136 (2017)
  doi:10.1007/JHEP01(2017)136
  [arXiv:1611.02598 [hep-th]].
\bibitem{Chernodub:2017ref}
  M.~N.~Chernodub and S.~Gongyo,
  Phys.\ Rev.\ D {\bf 95}, no. 9, 096006 (2017)
  doi:10.1103/PhysRevD.95.096006
  [arXiv:1702.08266 [hep-th]].

\bibitem{Bernard:1988db}
  V.~Bernard and U.~G.~Meissner,
  Nucl.\ Phys.\ A {\bf 489}, 647 (1988).
  doi:10.1016/0375-9474(88)90114-5
    \bibitem{Buballa:2003qv}
  M.~Buballa,
  Phys.\ Rept.\  {\bf 407}, 205 (2005)
  doi:10.1016/j.physrep.2004.11.004
  [hep-ph/0402234].
\bibitem{Kapusta}
  J.I. Kapusta, {\it Finite Temperature Field Theory},
  Cambridge University Press, Cambridge (1989).
\bibitem{Klevansky:1992qe}
  S.~P.~Klevansky,
  Rev.\ Mod.\ Phys.\  {\bf 64}, 649 (1992).
  doi:10.1103/RevModPhys.64.649

\bibitem{Bernard:1995hm}
  V.~Bernard, A.~H.~Blin, B.~Hiller, Y.~P.~Ivanov, A.~A.~Osipov and U.~G.~Meissner,
  Annals Phys.\  {\bf 249}, 499 (1996)
  doi:10.1006/aphy.1996.0081
  [hep-ph/9506309].


\bibitem{Schafer:1994nv}
  T.~Schafer, E.~V.~Shuryak and J.~J.~M.~Verbaarschot,
  Phys.\ Rev.\ D {\bf 51}, 1267 (1995)
  [hep-ph/9406210].


\bibitem{Schafer:1996wv}
  T.~Schafer and E.~V.~Shuryak,
  Rev.\ Mod.\ Phys.\  {\bf 70}, 323 (1998)
  doi:10.1103/RevModPhys.70.323
  [hep-ph/9610451].

\bibitem{Yu:2014sla}
  L.~Yu, H.~Liu and M.~Huang,
  Phys.\ Rev.\ D {\bf 90}, no. 7, 074009 (2014)
  doi:10.1103/PhysRevD.90.074009
  [arXiv:1404.6969 [hep-ph]].

\bibitem{Bratovic:2012qs}
  N.~M.~Bratovic, T.~Hatsuda and W.~Weise,
  Phys.\ Lett.\ B {\bf 719}, 131 (2013)
  doi:10.1016/j.physletb.2013.01.003
  [arXiv:1204.3788 [hep-ph]].

  \bibitem{Hell:2012da}
  T.~Hell, K.~Kashiwa and W.~Weise,
  J.\ Mod.\ Phys.\  {\bf 4}, 644 (2013)
  doi:10.4236/jmp.2013.45093
  [arXiv:1212.4017 [hep-ph]].

\bibitem{Li:2018ygx}
  Z.~Li, K.~Xu, X.~Wang and M.~Huang,
  arXiv:1801.09215 [hep-ph].


\bibitem{Luo:2017faz} 
  X.~Luo and N.~Xu,
  Nucl.\ Sci.\ Tech.\  {\bf 28}, no. 8, 112 (2017)
  doi:10.1007/s41365-017-0257-0
  [arXiv:1701.02105 [nucl-ex]].




\bibitem{Chao:2013qpa}
  J.~Chao, P.~Chu and M.~Huang,
  Phys.\ Rev.\ D {\bf 88}, 054009 (2013)
  doi:10.1103/PhysRevD.88.054009
  [arXiv:1305.1100 [hep-ph]].

\bibitem{Menezes:2009uc}
  D.~P.~Menezes, M.~Benghi Pinto, S.~S.~Avancini and C.~Providencia,
  Phys.\ Rev.\ C {\bf 80}, 065805 (2009).


\bibitem{Adamczyk:2013dal} 
  L.~Adamczyk {\it et al.} [STAR Collaboration],
  Phys.\ Rev.\ Lett.\  {\bf 112}, 032302 (2014)
  doi:10.1103/PhysRevLett.112.032302
  [arXiv:1309.5681 [nucl-ex]].


\bibitem{Aggarwal:2010wy} 
  M.~M.~Aggarwal {\it et al.} [STAR Collaboration],
  Phys.\ Rev.\ Lett.\  {\bf 105}, 022302 (2010)
  doi:10.1103/PhysRevLett.105.022302
  [arXiv:1004.4959 [nucl-ex]].
\bibitem{Bazavov:2017dus} 
  A.~Bazavov {\it et al.},
  Phys.\ Rev.\ D {\bf 95}, no. 5, 054504 (2017)
  doi:10.1103/PhysRevD.95.054504
  [arXiv:1701.04325 [hep-lat]].
   \end{thebibliography}
\end{document}